\documentclass[reprint,
showpacs,
showkeys,
floatfix,
amsmath,amssymb,
 aps,
pra,
]{revtex4-1}
\usepackage{graphicx}
\usepackage{dcolumn}
\usepackage{bm}
\usepackage{hyperref}
\usepackage[USenglish]{babel}
\usepackage{physics}
\usepackage[shortlabels]{enumitem}
\usepackage{amsmath}

\DeclareMathOperator{\sgn}{sgn}

\begin{document}

\preprint{APS/123-QED}

\title{Analytical description of spontaneous emission of light at the optical event horizon}

\author{Maxime J Jacquet}
 \affiliation{Vienna Center for Quantum Science and Technology, Faculty of Physics, University of Vienna, Boltzmanngasse 5, Vienna A-1090, Austria.}
 \email{maxime.jacquet@univie.ac.at}
\author{Friedrich K\"{o}nig}%
 \email{fewk@st-andrews.ac.uk}
\affiliation{School of Physics and Astronomy, SUPA, University of St. Andrews, North Haugh, St. Andrews, KY16 9SS, United Kingdom}%
\date{\today}

\begin{abstract}
Quantum fluctuations in curved space-time cause the emission of particles.  In order to understand how they may be detected in a laboratory experiment, we consider a moving refractive index perturbation in an optical medium, which exhibits optical event horizons. Based on the field theory in curved space-time we formulate an analytical method to calculate the scattering matrix that completely describes mode coupling leading to the emission of photon pairs in various configurations. We then quantify the spectrally resolved photon number correlations. Moreover, we apply our method in a case study, in which we consider a moving refractive index step in bulk fused silica. We calculate key observables in the moving frame as well as in the laboratory frame, such as the emission spectrum and the spectrally resolved quantum correlations of the photon number. We observe significant spectral correlations between modes of opposite norm, evidence of their vacuum origin. We find that emission from horizons is characterized by an increased photon flux, a signature spectral shape as well as a correlation with the partner photon mode approaching unity. These methods and findings pave the way to the observation of particles from the event horizon in dispersive systems.  
\end{abstract}



\maketitle


A number of classical and semi-classical effects of black-hole physics can be reproduced in the laboratory.
In particular, it is possible to create event horizons for waves in media \cite{unruh_experimental_1981}.
These horizons  scatter waves and are predicted to  spontaneously emit quanta by the Hawking effect \cite{hawking_black_1974,hawking_particle_1975}.
Recently, experiments in many different `fluid' systems, such as Bose-Einstein condensates \cite{lahav_realization_2010,steinhauer_observation_2014,steinhauer_observation_2016},  water \cite{rousseaux_observation_2008,weinfurtner_measurement_2011,Rousseaux_PRL_2016,torres_rotational_2017,euve_wormholes_2017} and  polariton microcavities \cite{nguyen_acoustic_2015} have demonstrated horizons and studied the behaviour of waves in their vicinity.
An unequivocal observation of the Hawking effect in these fluid systems remains to be performed.

For light, it is also possible to create an event horizon \cite{leonhardt_relativistic_2000,leonhardt_laboratory_2002,schutzhold_EMW_2005,gorbach_light_2007,philbin_fiber-optical_2008,faccio_analogue_2010}.
Experimentally, this can be done by changing the speed of light with light itself \cite{philbin_fiber-optical_2008, Hill_soliton-trapping_2009}.
A short and intense laser pulse locally raises the refractive index of a medium by the Kerr effect: under the pulse, waves will be slowed.
Hence, the profile of refractive index created by the propagating pulse effectively sets the curvature of the spacetime on which waves propagate.
If light under the pulse is slowed below the pulse speed, the pulse moves superluminally, and two horizons are formed at the boundaries between sub- and superluminal propagation: light cannot enter the back of the pulse or is captured falling into the front of the pulse.
In analogy with the metric of spacetime in the vicinity of a black hole \cite{unruh_experimental_1981,visser_acoustic_1998,p._lisle_river_2006,barcelo_hawking-like_2006}, the back and the front of the pulse thus act as a white-hole or a black-hole event horizon, respectively \cite{philbin_fiber-optical_2008}.

As in the case of the astrophysical black hole horizon \cite{hawking_particle_1975}, spontaneous emission in analogue systems results from the mixing of field modes of positive and negative Klein-Gordon norm at the horizon \cite{unruh_experimental_1981}.
The paired emission allows to identify and characterize the effect.
This has been extensively studied for fluid experiments \cite{finazzi_entangled_2014,de_nova_violation_2014,de_nova_time-dependent_2015,boiron_quantum_2015}, and is considered an unmistakable signature of the Hawking effect at the event horizon \cite{schutzhold-unruh-comment-2011,unruh_has_2014,finke_observation_2016}.

Compared to their fluid-based siblings, optical horizons are less well understood, because of the complicated dispersion in connection with the breakdown of the JWKB approximation at the horizon. 
And yet, they are important due to the high emission temperature \cite{philbin_fiber-optical_2008} and the ability to directly manipulate and utilize individual entangled pairs of quanta. They offer directly observable spectra and close to maximal entanglement. First spontaneous spectra  have been numerically computed in \cite{petev_blackbody_2013, liberati_quantum_2012, belgiorno_perturbative_2014, wang_optical_2015}, but cannot describe the role of modes at the event horizon and omit negative norm waves. 
A number of analytical studies based on a Taylor-expansion description of the refractive index \cite{robertson_hawking_2011,robertson_integral_2014_I,robertson_integral_2014} have been proposed, but the models are limited to single branches of the dispersion relation. 
On the other hand, the Hopfield model \cite{hopfield_theory_1958} realistically  describes the refractive index of dispersive dielectrics \cite{schutzhold_dielectricbh_2002} and a fully quantized analytical model can be constructed for spontaneous emission at a moving, step-like, refractive index front (RIF) in a nonlinear dielectric \cite{finazzi_quantum_2013}.
Even smooth RIF profiles  can be computed but the emission spectra or photon number correlations are difficult to compute with realistic dispersion relations or variable pulse shapes \cite{belgiorno_hawking_2015,linder_derivation_2016}.
Conversely, the model \cite{finazzi_quantum_2013}, used in \cite{finazzi_spontaneous_2014} and \cite{jacquet_quantum_2015}, allows to readily and directly compute spectra of spontaneous emission in the medium rest frame of the measurement.
In particular, it was realised that a moving RIF acts simultaneously as a black hole-, white hole- and horizonless emitter \cite{jacquet_quantum_2015}.
These kinematic aspects are important because they might shed light on experimental results \cite{weinfurtner_measurement_2011,Rousseaux_proceedings_2016,coutant_subcritical_2016}.
Furthermore, the understanding of the various possible configurations for waves at the RIF led to the first computation of a laboratory-frame spectrum of spontaneous emission featuring both positive- and negative-norm waves \cite{jacquet_quantum_2015}. In oder to support an experimental observation, a formalism to calculate laboratory spectra and their photon number correlations has to be developed.

In this paper, we show how to analytically calculate the scattering matrix in a realistic dispersive dielectric medium described by the canonical model \cite{finazzi_quantum_2013}. The scattering matrix completely describes the mode coupling classically and quantum-mechanically.
With the example of a step-like RIF, we use the analytical approach to identify five distinct kinematic configurations of mode interaction at the analogue horizon.
The step-RIF is not only the simplest geometry generating optical horizons, but also experimentally realistic \cite{gaafar_plasma-mirror-horizon_2017,ciret_observation_2016,kanakis_enabling_2016}.
It has provided reliable results when the medium parameters do not vary much across the step \cite{finazzi_hawking_2012}\footnote{This is valid for small index changes \cite{Jacquet_PhDThesis_2017}. In \cite{philbin_fiber-optical_2008,choudhary_efficient_2012} or \cite{Jacquet_PhDThesis_2017}, the index change is  $\delta n\sim10^{-6}$.}, and more complex and smooth index profiles can be constructed by a piecewise constant approximation.
We demonstrate the calculation of key observable quantities in the medium rest frame, \textit{i.e.}, the photon emission spectral density and the spectrally-resolved photon-number correlations, as they are recorded with a variable bandwidth detector.
We obtain spectra that feature strong peaks and associated pairwise emission from horizons on a background of broadband horizonless emission.
Both the horizon-emission peaks and the strong photon number correlations are principal identifiers of spontaneous emission from the vacuum at the horizon by the Hawking effect. Our method allows to consider all possible configurations of frequency, RIF-height and RIF-velocity to analytically compute the scattering matrix.

In section \ref{sec:FTIDD}, we briefly restate the field theory of light in a dispersive medium.
We review all possible kinematic configurations for waves at the RIF, construct the inhomogeneous plane-wave solutions and introduce their scattering matrix. We also quantise the inhomogeneous solutions. Section \ref{sec:N} derives the photon flux in different frames as well as the spectral correlations detected with variable bandwidth detectors in terms of the scattering matrix. 
Section \ref{sec:analyticsS} covers the field matching conditions at the RIF to derive the scattering matrix.  In section \ref{sec:numerics} we exemplify the use of our method by a case study. We calculate key observable quantities, the spectral density of spontaneous emission and the  spectral correlations, in the rest frame of the medium and discuss their dependence on the RIF velocity. We also briefly discuss desired medium properties.

\section{\label{sec:FTIDD}Light in an inhomogeneous dispersive dielectric}

Following \cite{schutzhold_dielectricbh_2002,finazzi_quantum_2013} and \cite{jacquet_quantum_2015}, we describe the interactions of light with an inhomogeneous and transparent dielectric by a microscopic model based on the Hopfield model \cite{hopfield_theory_1958}.
We consider one-dimensional scalar electromagnetic fields and operate at frequencies sufficiently far from the medium resonances to neglect absorption.
The medium consists of polarisable molecules --- oscillators with eigenfrequencies (resonant frequency) $\Omega_i$ and elastic constants $\kappa_i^{-1}$ ($i=1,2,3$).
Since the wavelength of light is large compared to the molecular scale, we consider the dielectric in the continuum limit and describe the electric dipole displacement by the massive scalar field $P_i$.
The electromagnetic field (a massless scalar field) is described by the vector potential $\vec{A}(X,T)$ via $\vec{E}=-\partial_T \vec{A}$ in temporal gauge, where $X$ and $T$ are space and time in the laboratory frame.
The refractive index $n$ of most materials is well described by a medium featuring three resonances.

\begin{figure}[t]
\centering
\includegraphics[width=.45\columnwidth]{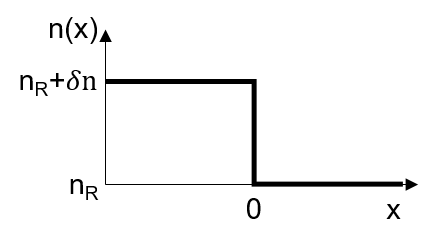}
\caption[Sketch of the refractive index front (RIF) in the moving frame]{Sketch of the refractive index front (RIF) in the moving frame: there are two homogeneous index regions on the left and on the right of a dielectric boundary of height $\delta n$.}
\label{fig:RIF}
\end{figure}

Our study is based on the consideration of the simple step-like geometry of a RIF, that propagates at constant speed $u$ in the positive $X$-direction in the laboratory frame. 
The RIF is shown in Fig.\ref{fig:RIF} in \textit{co-moving frame} coordinates $x$ and $t$. We locate the boundary of the RIF at $x=0$. We focus entirely on the index change induced by the RIF, neglecting phonon interactions and phasematched optical nonlinearities such as four-wave mixing.
In both homogeneous regions ($x \gtrless 0$), the interaction of the electromagnetic field with the three polarization fields of the medium is described by the Lagrangian density \cite{finazzi_quantum_2013,jacquet_quantum_2015,hopfield_theory_1958,fano_atomic_1956}
\begin{equation}
\label{eq:LagrangianHopfieldMF}
\begin{split}
\mathcal{L}_{MF}&=\frac{(\partial_t A)^2}{8 \pi c^2}-\frac{(\partial_x A)^2}{8 \pi}\\
&\ \ \ +\sum_{i=1}^3\left(\frac{\gamma^2(\partial_t P_i-u\partial_x P_i)^2}{2\kappa_i\Omega_i^2}\right.\\
&\ \ \ \left.-\frac{P_i^2}{2\kappa_i}+\frac{A\gamma}{c}(\partial_t P_i-u\partial_x P_i)\right),
\end{split}
\end{equation}
where $(\kappa_i\Omega_i^2)^{-1}$ is the inertia of oscillator $P_i$  and $\gamma=\left(1-u^2/c^2\right)^{-1/2}$.
The term linear in $A$ in Eq.\eqref{eq:LagrangianHopfieldMF} describes the coupling between the fields.
The Lagrangian density accounts for the free space and medium contributions to the field through the first two terms and the sum, respectively.
Dispersion enters as a time dependence of the addends of the summation.
By the principle of least action, we obtain the Hamiltonian density by varying the Lagrangian density \eqref{eq:LagrangianHopfieldMF} with respect to the canonical momentum densities of light and the polarisation fields.
From the Hamiltonian density follow the Hamilton equations, the equations of motion for the fields \cite{cohen-tannoudji_photons_2004,finazzi_quantum_2013}:
\begin{eqnarray}
\label{eq:fieldmotioneqMFFT}
\label{eq:fieldmotioneqMFFT1}
\dot{A}&=&4\pi c^2\Pi_A\\
\label{eq:fieldmotioneqMFFT2}
\dot{P}_i&=&
\frac{\kappa_i\Omega_i^2}{\gamma^2}
\left(\Pi_{P_i}-A\frac{\gamma}{c}\right)+uP_i^{'}\\
\label{eq:fieldmotioneqMFFT3}
\dot{\Pi}_A&=&\frac{A^{''}}{4\pi}+\sum_{i=1}^3\left(\frac{\kappa_i\Omega_i^2}{\gamma^2}
\left(\Pi_{P_i}-A\frac{\gamma}{c}\right)\right)\\
\label{eq:fieldmotioneqMFFT4}
\dot{\Pi}_{P_i}&=&-\frac{P_i}{\kappa_i}+u\Pi_{P_i}^{'},
\end{eqnarray}
where derivatives are with respect to $(x, t)$ and $\Pi$'s denote the canonical conjugate fields.
We complexify the massive field obtained from the action of \eqref{eq:LagrangianHopfieldMF} by identifying harmonic plane wave solutions  to \eqref{eq:fieldmotioneqMFFT} of the form
\begin{equation}
\label{eq:PWsolsMF}
\vec{V}(x, t)=\vec{\bar{V}}({\omega})\,e^{ik x-i\omega t},
\end{equation}
where $\vec{V}$ is the eight-dimensional field vector $\vec{V}=(A\ P_1\ P_2\ P_3\ \Pi_A\ \Pi_{P_1}\ \Pi_{P_2}\ \Pi_{P_3})^T$.
We go to Fourier space by $\partial_t \leftrightarrow -i\omega$ and $\partial_x \leftrightarrow ik$, where $k$ and $\omega$ are, respectively, the wavenumber and frequency in the moving frame. We obtain the generic Sellmeier dispersion relation of bulk transparent dielectrics: 
\begin{equation}
\label{eq:SellmDispRelMF}
c^2k^2=\omega^2+\sum_{i=1}^3\frac{4\pi\kappa_i\gamma^2\left(\omega+uk\right)^2}{1-\frac{\gamma^2\left(\omega+uk\right)^2}{\Omega_i^2}}.
\end{equation}
This dispersion relation describes the index in each of the homogenous regions on either side of the RIF (see Fig.\ref{fig:RIF}). We can denote a harmonic field vector as  $\vec{V}^{\alpha}({\omega})$, where $\alpha$ indicates a particular solution of  \eqref{eq:SellmDispRelMF} for $\omega$, i.e. a mode. 

By construction, the Lagrangian \eqref{eq:LagrangianHopfieldMF} is invariant under global phase shifts of the dynamic fields.
This continuous symmetry implies a conserved Noether current \cite{Jacquet_PhDThesis_2017, linder_derivation_2016}.
As a result,  the Klein-Gordon product
\begin{equation}
\label{eq:scalarproduct}
\left\langle \vec{V}_1,\vec{V}_2 \right\rangle =\frac{i}{\hbar}\int dx \, \vec{V}_1^\dagger(x,t)\,\left(\begin{array}{cc}0&I_4\\-I_4&0\end{array}\right)\, \vec{V}_2(x,t).
\end{equation}
 is conserved, and so is the induced norm. Here $I_4$ is the $4\times4$ identity matrix and the Planck constant  prefactor was inserted for normalisation.
It can be shown that the induced norm of a positive (negative) laboratory frequency $\Omega$ 
field is also positive (negative) \cite{finazzi_quantum_2013,rubino_negative-frequency_2012,mclenaghan_compression_2014}. This is different to fluid systems, where the sign of the norm is equal to that of the wave number. In optical systems, the Hawking effect takes place in correlated photons of positive and negative frequency, not wavenumber.
As a result, waves of positive frequency $\omega$ in the moving frame can have either sign of the norm.

We orthonormalise the field vectors $\vec{V}^{\alpha}({\omega})$  using the scalar product \eqref{eq:scalarproduct} and the condition \cite{finazzi_quantum_2013}:
\begin{equation}
\label{eq:normalisationcondition}
\left\langle \vec{V}({\omega})^{\alpha_1}, \vec{V}({\omega'})^{\alpha_2}\right\rangle=\sgn(\Omega)\,\delta_{\alpha_1\alpha_2}\, \delta(\omega-\omega').
\end{equation}
Here $\sgn$ is the sign function that determines whether the mode $\alpha_1$ has positive or negative norm. Moreover, by Lorentz transform $\Omega = \gamma (\omega+uk^{\alpha})$ and $K^{\alpha} = \gamma (k^{\alpha}+u/c^2\,\omega)$.

We now describe the non-uniform medium.
The index in each  homogeneous region is described by the dispersion relation \eqref{eq:SellmDispRelMF}, with dispersion parameters $\kappa_{i,R}$ ($\kappa_{i,L}$) and $\Omega_{i,R}$ ($\Omega_{i,L}$) in the right (left) region.
The index distribution in the moving frame is (Fig.\ref{fig:RIF}):
\begin{equation}
\label{eq:RIFheight}
n(x)=n_L \,\theta \left( -x \right) +n_R \,\theta \left( x \right) =n_R+\delta n\, \theta \left( -x \right).
\end{equation}
$\theta$ is the Heaviside step function and $n_R$ ($n_L$) is the index on the right (left) side.
In an extension of the oscillator model by P. Drude and H. A.  Lorentz, $\mu$ parametrises the change of dispersion constants that leads to the  index change $\delta n$ \cite{schubert_nonlinear_1986}:
\begin{equation}
\label{eq:nlindex}
 \kappa_{i L}=\mu\kappa_{i R} \quad \quad \Omega_{i L}^2=\mu^{-1}\Omega_{i R}^2.
\end{equation} 
For small index changes it follows from \eqref{eq:SellmDispRelMF} that $\mu\approx1+2 (n_R -n_R^{-1})^{-1} \, \delta n$.

Harmonic wave solutions of frequency $\omega$ have a  propagation constant $k$ given by  \eqref{eq:SellmDispRelMF}, which is an eighth order polynomial, and thus eight wavenumbers $k^{\alpha}$ form the modes of the field $\vec{V}$ with degenerate energy $\hbar \omega$.
On either side of the RIF, there are either eight propagating modes or six propagating modes and two exponentially growing and decaying modes, respectively, characterized by complex $\omega$ and $k$ (cf. Fig.\ref{fig:labdisprel}). Here we discuss optical event horizons that are realised for waves in the optical frequency branch. Although we do not neglect scattering contributions of the non-optical modes, we focus on the kinematics and spectral properties of the optical modes.

\begin{figure}[h]
\centering
\includegraphics[width=.9\columnwidth]{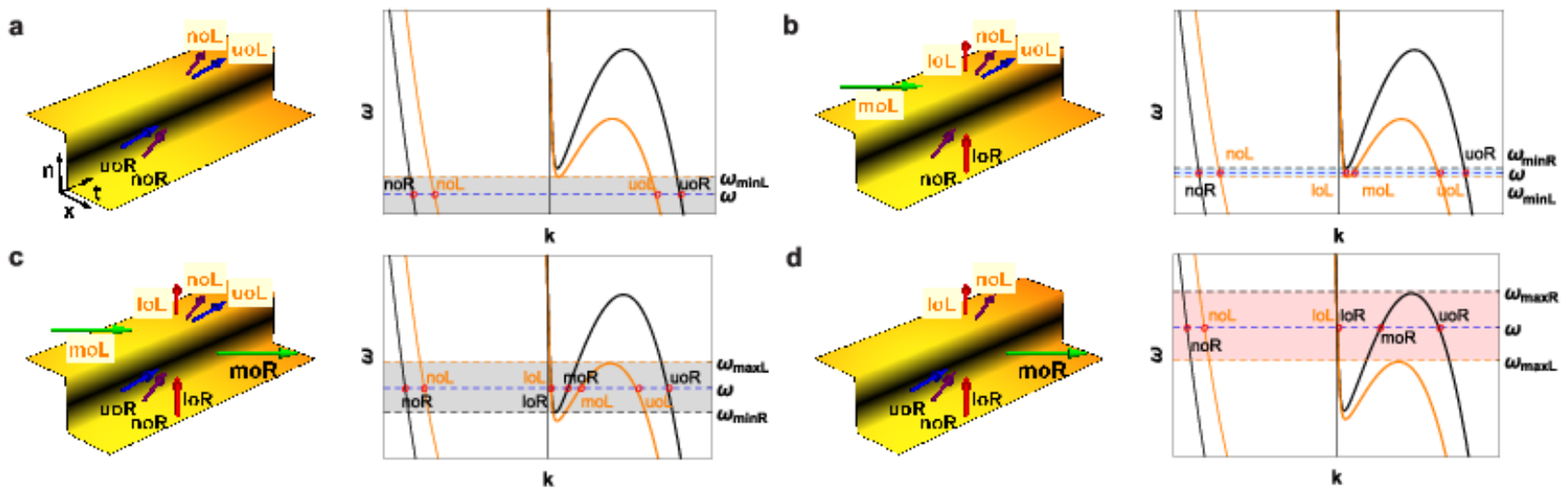}
\caption[Mode configurations at the RIF]{Mode configurations for different comoving frequencies $\omega$ (a-e). Shown on the left are schematics of propagating modes on either side of the RIF.  The arrows indicate the direction of group velocity for each mode. Modes noL and noR are the only negative-norm optical modes on the left and right of the step, respectively. On the right the dispersion diagram in the moving frame is displayed. The black (orange) line corresponds to the low (high) index region at $x>0$ ($x<0$). The blue dashed line at frequency $\omega$ intersects with the dispersion curve and identifies single-frequency modes.  The step acts as a black hole horizon over the orange-shaded frequency interval in (d), and as a white hole-like horizon over the blue-shaded interval in (b).}
\label{fig:intervals}
\end{figure}

In Fig.\ref{fig:intervals} we lay out all possible configurations of optical modes at either side of the RIF.
The dispersion diagram shows the optical branch for $x>0$ ($x<0$) in black (orange). The negative norm part of the branch shows on the left in the diagram (thin lines) and the positive norm part lies on the right (thick) lines. 
The number of propagating modes depends on $\omega$: for all $\omega$, there is one negative-norm mode and either  one or three positive-norm modes. 
We refer to frequency intervals of  $\omega$  on either side of the RIF with three positive norm modes as \textit{subluminal intervals} (SLIs): $\left[\omega_{min L},\omega_{max L}\right]$ and $\left[\omega_{min R},\omega_{max R}\right]$. 
Inside a SLI, one of the four mode solutions has a positive group velocity $\frac{\partial\omega}{\partial k}$ in the moving frame.
On the right-hand-side of the RIF ($x \geq 0$), this  is the unique mode in which light may propagate away from the RIF.
We call this mode `mid-frequency optical on the right', `$moR$'.
The other three modes have negative group velocities and move into the boundary from the right.
There is the low-frequency optical mode \textit{loR}, upper frequency optical mode \textit{uoR}, and the negative frequency optical mode \textit{noR}. 
On either side we can order the modes by the wave number $k$ and obtain $k^{\,noR/L}({\omega})\leq k^{\,loR/L}({\omega})\leq k^{\,moR/L}({\omega})\leq k^{\,uoR/L}({\omega})$ (see Fig.\ref{fig:intervals} (c)).
Beyond the SLI, \textit{i.e.} for $\omega\leq \omega_{min}$ or $\omega\geq \omega_{max}$, only two propagating modes remain.
Thus in growing order of $\omega$ we find the following mode configurations at the step, as presented in Fig.\ref{fig:intervals} \footnote{For a more detailed discussion of the mode configurations see \cite{Jacquet_PhDThesis_2017}}:
\begin{enumerate}[(a)]
\item $\omega<\omega_{minL}$, Fig.\ref{fig:intervals} (a).  Two optical propagating modes (\textit{uoL/R},\textit{noL/R}) exist, with negative group velocities in the moving frame on either side.
No optical horizon exists.
\item $\omega_{minL}<\omega<\omega_{minR}$, Fig.\ref{fig:intervals} (b). On the left of the interface, there exist four optical propagating modes (\textit{loL, moL,uoL}, \textit{noL}) whilst only modes \textit{uoR} and \textit{noR} exist on the right.
Mode \textit{moL} on the left is the only mode with positive group velocity.
Light can propagate into the boundary from the left, but cannot proceed further to the right.
The interface acts as a white hole horizon to light.
\item $\omega_{minR}<\omega<\omega_{maxL}$, Fig.\ref{fig:intervals} (c). Four propagating modes (\textit{noL/R, loL/R, moL/R}, \textit{uoL/R}) exist on either side of the interface.
Mode \textit{moL}(\textit{R}) is the only mode with positive group velocity on the left (right) of the RIF.
The RIF is not a one-way door and thus no horizons exist.
\item $\omega_{maxL}<\omega<\omega_{maxR}$, Fig.\ref{fig:intervals} (d). Light on the right can move in either direction, but on the left of the interface both modes have negative group velocity. 
Light experiences a black-hole horizon at the RIF.
\item $\omega>\omega_{maxR}$, Fig.\ref{fig:intervals} (e). Similarly to (a), two propagating modes (\textit{noL/R, loL/R}) exist on either side of the interface with negative group velocities.
No optical horizon exists.
\end{enumerate}
In configurations (b) and (d) a subluminal region is paired with a superluminal region, creating a horizon. This is in analogy to the super- (sub-)luminal space flow in the interior region of a black- (white-)  hole and the subluminal flow outside \cite{visser_analogue_2002,p._lisle_river_2006}. Thus the  index step plays the role of the disturbance in the space-time geometry. 

Now that we have discussed all possible mode configurations, we proceed to construct modes of the inhomogeneous system, the  global modes (GMs).
We connect the plane wave solutions $\vec{V}(x, t)$,  the local modes (LMs), for the homogeneous medium at the  index boundary at $x=0$. 
We show in Appendix \ref{app:matchingconditions}, that all fields $A, P_i, \Pi_A$, and $\Pi_{P_i}$ are continuous at the boundary. All derivatives  of all fields in space and time are also continuous, except for $\Pi_{P_i}$. We refer to these relations as the `matching conditions'.

We construct GMs $\mathcal{\vec{V}}$ as
\begin{equation}
\label{eq:GMdef}
\vec{\mathcal{V}}(x, t)=\sum_\alpha L^\alpha\, \vec{V}_L^\alpha(x, t)\,\theta(-x)+\sum_\alpha R^\alpha \,\vec{V}_R^\alpha(x, t)\,\theta(x),
\end{equation}
where $L^{\,\alpha}$ ($R^{\,\alpha}$) are coefficients of the eight modes $\alpha$ on the left (right) side of the RIF.
Half of the 16 coefficients in \eqref{eq:GMdef} are constrained by the matching conditions, leaving eight independent global modes.

\begin{figure}[h]
\centering
\includegraphics[width=.25\textwidth]{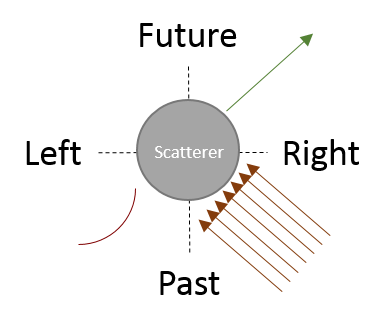}
\caption[Mode composition of the \textit{out} GM \textit{moR}]{Mode composition of the global \textit{out} mode \textit{moR} in a spacetime diagram. There is a unique mode that propagates away from the RIF to the right (green arrow). In the past, 7 oscillatory-modes propagate toward the RIF from the right and there is one complex decaying mode on the left of the scatterer (RIF).}\label{fig:modedecompomoR}
\end{figure}

We construct 16 particular GMs from the 16 LMs with their group velocity either towards (\textit{in}) or away from (\textit{out}) the RIF \cite{macher_black/white_2009}, irrespective of whether the LM is on the right or on the left.
Half of the GMs emerge from a defining \textit{in} LM $\alpha$, forming global \textit{in} modes $\vec{\mathcal{V}}^{in\,\alpha}$. 
The others are global \textit{out} modes $\vec{\mathcal{V}}^{out\,\alpha}$ where $\alpha$ is an \textit{out} LM. Where there exist  8 propagating LMs on either side of this interface, we find 8 \textit{in} and 8 \textit{out} GMs. We arrange these in decreasing order of laboratory-frame frequency $\Omega$ to allow for a consistent treatment of the matching conditions. If, on either side, two modes are complex, they will have complex conjugate wave numbers and frequencies. In this case the unbounded LM will define an unphysical GM, without other LMs on that side, which serves doubly as \textit{in} as well as \textit{out} mode.
The LMs are complete solutions in the homogeneous regions, i.e. the sets of $\vec{\mathcal{V}}^{in}$ and $\vec{\mathcal{V}}^{out}$ modes are two basis sets.
Let us consider the example of the black hole-like case (mode configuration (d)). There is a unique \textit{out} GM, mode \textit{moR}, that allows for light to propagate away from the interface into the low index region.
Its mode decomposition is shown in a  spacetime diagram in Fig.\ref{fig:modedecompomoR}: it is a linear combination of 7 oscillatory LMs, in the right region, that have negative group-velocity, a non-oscillatory LM on the left and a unique mode that has positive group-velocity in the right region.

We use the S-matrix formalism to relate incoming and outgoing fields. The scattering matrix $S$ transforms the \textit{out} field modes to the \textit{in} modes:
\begin{equation}
\label{eq:Smatrixdef}
\vec{\mathcal{V}}^{in\,\alpha}=\sum_\beta \vec{\mathcal{V}}^{out\,\beta}S_{\beta\,\alpha}.
\end{equation}
Forming a matrix $\mathcal{V}^{\mathrm{in}}$ ($\mathcal{V}^{\mathrm{out}}$) from the \textit{in} (\textit{out}) basis set, we describe the basis change as:
\begin{equation}
\begin{split}
\label{eq:fieldtransfo}
{\mathcal{V}}^\mathrm{in}&=\left(\vec{\mathcal{V}}^{\mathrm{in}\,\alpha_1}\ \vec{\mathcal{V}}^{\mathrm{in}\,\alpha_2}\ ... \ \vec{\mathcal{V}}^{\mathrm{in}\,\alpha_8}\right) \\ &=\left(\vec{\mathcal{V}}^{\mathrm{out}\,\alpha_1}\ \vec{\mathcal{V}}^{\mathrm{out}\,\alpha_2}\ ... \ \vec{\mathcal{V}}^{\mathrm{out}\,\alpha_8}\right)\, S={\mathcal{V}}^\mathrm{out}\,S.
\end{split}
\end{equation}
The spontaneous photon creation occurs because the quantum vacuum is basis dependent. Hence the spontaneous emission and all mode conversion follows from $S$.

We proceed with the canonical quantisation formalism  introduced in \cite{fano_atomic_1956}, developed in the 1990s in \cite{huttner_canonical_1991,huttner_dispersion_1992,matloob_electromagnetic_1995,barnett_quantum_1995,santos_electromagnetic-field_1995,barnett_spontaneous_1992}, and used in \cite{finazzi_quantum_2013,linder_derivation_2016,belgiorno_hawking_2015} and \cite{jacquet_quantum_2015,Jacquet_PhDThesis_2017} for the global modes.
We postulate the equivalent of the standard equal-time commutation relations on the fields $A$ and $P_i$:
\begin{equation}
\label{eq:commutA}
\left[A(x),\Pi_A(x')\right]=i\hbar\,\delta(x-x'),
\end{equation}
\begin{equation}
\label{eq:commutPi}
\left[P_i(x),\Pi_{P_j}(x')\right]=i\hbar\,\delta_{ij}\,\delta(x-x').
\end{equation}
We quantise the GMs by writing the global field $\vec{\hat{\mathcal{V}}}$ in the basis of global \textit{in} modes:
\begin{multline}
\label{eq:GMinLMinbasis}
\vec{\hat{\mathcal{V}}}=\int\limits_0^\infty\mathrm{d\omega}\left( \sum_{\alpha\in P}\vec{\mathcal{V}}^{in\,\alpha}({\omega})\,\hat{a}^{in\,\alpha}({\omega}) \right. \\   + \left. \sum_{\alpha\in N}\vec{\mathcal{V}}^{in\,\alpha}({\omega})\,\hat{a}^{in\,\alpha\dagger}({\omega}) \right)+ \mathrm{H.c.},
\end{multline}
where P (N) denotes the set of all positive (negative) norm global modes \footnote{Unphysical modes can be associated with positive norm.}.
The expansion \eqref{eq:GMinLMinbasis} for \textit{in} (and its counterpart for \textit{out} modes) defines the annihilation operators  $\hat{a}^\alpha({\omega})$ and the creation operators $\hat{a}^{\alpha\dagger}({\omega})$ for each global mode $\alpha$, as well as the transformation between \textit{in} and \textit{out} creation and annihilation operators.
Hence, let $\vec{\hat{A}}^{in}$ be the column vector containing all the annihilation and creation operators for positive- and negative-norm global \textit{in} modes, respectively, and $\vec{\hat{A}}^{out}$ the corresponding vector for the \textit{out} modes. Then the transformation of operators follows from \eqref{eq:Smatrixdef} and \eqref{eq:GMinLMinbasis} as \cite{jacquet_quantum_2015}:
\begin{equation}
\label{eq:operatortransfo}
\vec{\hat{A}}^{out}=S\vec{\hat{A}}^{in}.
\end{equation}

  With the scalar product \eqref{eq:scalarproduct} and the mode expansion \eqref{eq:GMinLMinbasis}  we can derive explicit expressions for the annihilation and creation operators and show that their commutator follows \cite{finazzi_quantum_2013}:
\begin{equation}
\label{eq:commutator}
\left[\hat{a}^{\alpha}({\omega}),\hat{a}^{\alpha'\dagger}({\omega'})\right]= \delta_{\alpha \alpha'}\delta(\omega-\omega'),
\end{equation}
where the first $\delta$ is the Kronecker-delta. This relation holds for \textit{in} and \textit{out} operators. The commutator confirms that the global modes defined here are independent Bosonic modes.

\section{Spectral densities and photon-number correlations}\label{sec:N}

The aim of field theory in section \ref{sec:FTIDD} is to calculate the observable spectral density and spectral correlations as observed in the laboratory frame. In this section we calculate the photon flux, photon number and the photon number correlations in terms of the scattering matrix $S$. 

We calculate the spontaneous emission in \textit{out} modes, i.e. the  \textit{in} modes are in the vacuum state. The \textit{out} GMs have to have a \textit{positive} laboratory-frame group velocity to reach the detector. 
\begin{figure}[t]
\centering
\includegraphics[width=\columnwidth]{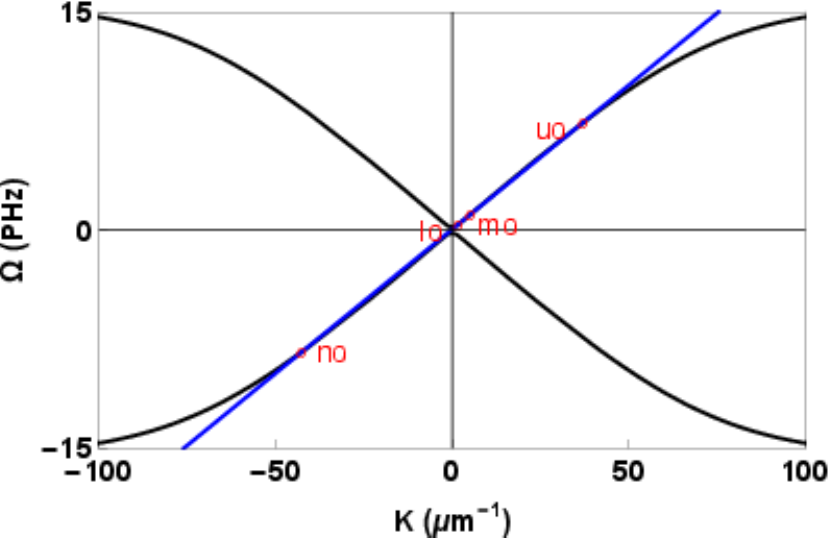}
\caption{Optical branch of the dispersion in bulk fused silica in the laboratory frame. The positive- and negative-frequency branches are shown in black. At points of intersection with a contour of $\omega$ (blue straight line), plane wave mode solutions are named (red circles).\label{fig:labdisprel}}
\end{figure}
As can be seen in Fig.\ref{fig:labdisprel}, modes \textit{no} and \textit{uo} have positive group velocities smaller than the RIF velocity $u$. This also applies to mode \textit{lo}, except that it has a negative group velocity if $K$ is negative. These modes are \textit{out} GMs on the left.  In the low-index region on the right, the only \textit{out} GM, i.e. propagating \textit{away} from the RIF ahead of it, is mode \textit{moR}.
Thus we expect to detect contributions from GMs \textit{noL}, \textit{uoL}, \textit{loL} and \textit{moR}.

Because the system is in a stationary state, the photon flux for an \textit{out} mode $\alpha$ is given by the integral over the frequency correlation \cite{loudon_2000}:
\begin{equation}
\begin{split}
\label{eq:fluxop}
\phi^\alpha(\omega)&=\left \langle 0^{\mathrm{in}}|\hat{\phi}^\alpha(\omega)|0^{\mathrm{in}} \right \rangle \\ &= \int\limits_0^\infty \frac{d\omega'}{2 \pi} \,\left \langle 0^{\mathrm{in}} |\hat{a}^{\mathrm{out}\, \alpha \dagger}({\omega}) \hat{a}^{\mathrm{out}\, \alpha}({\omega'})|0^{\mathrm{in}} \right \rangle
\end{split}
\end{equation}
The flux $\phi(\omega)$ is the dimensionless number of photons per unit time and unit angular frequency at $\omega$ in the moving frame. The \textit{out}-mode operators in the  frequency correlation are replaced using \eqref{eq:operatortransfo} by \textit{in}-operators, which act directly on the \textit{in} vacuum state, see appendix \ref{app:corrfunctions}. Using eq. \eqref{eq:2ndmom1}, the result for the photon flux is:

\begin{equation}
\label{eq:flux}
\phi^\alpha(\omega)=\frac{1}{2 \pi}\sum_{\beta \notin \{\alpha\}}  {\left|S_{\alpha \beta}(\omega)\right|^2}.
\end{equation}
Here $\{\alpha\}$ is the set of modes which have a norm equal to $\alpha$. The photon flux results from the scattering of \textit{in} modes into \textit{out} modes of opposite norm.
The number of emitted photons depends solely on the scattering matrix, bandwidth and interaction time.

A second important observable is  the photon-number correlation across the spectrum. This is an essential signature of the expected entanglement between photons of different wavelengths.
In the moving frame, the photon flux density $\phi^\alpha({\omega})$ in mode $\alpha$ at frequency $\omega$ is the dimensionless number of photons per unit time $\delta\tau$ and unit bandwidth $\delta\omega$, cf. \eqref{eq:flux}. The total photon number is obtained by integrating over all frequencies and time. To obtain the photon number $\langle \hat{N} \rangle$ over a limited `detected' frequency interval in this stationary mode conversion process, the field operators have to be constrained to this bandwidth. The photon number operator is then given by:
\begin{equation}
\label{eq:hatN}
\hat{N}^\alpha_t=\tau \bigg(\int\limits_0^\infty \!\! \frac{d\omega}{\sqrt{2 \pi}} t(\omega) \hat{a}^{\mathrm{out}\,\alpha}(\omega)\bigg)^{\dagger} \bigg(\int\limits_0^\infty \!\! \frac{d\omega}{\sqrt{2 \pi}} \\ t(\omega) \hat{a}^{\mathrm{out}\,\alpha}(\omega)\bigg).
\end{equation}

Here $\tau$ is the interaction time in the \textit{moving} frame and $t(\omega)$ is a spectral filter function, which takes the value of 1 if the frequency is in the detected interval and 0 otherwise.
Correlations in photon number between modes $\alpha$ and $\alpha'$ are found by calculating the normally ordered covariance of the photon numbers $\hat{N}_1^\alpha$ and $\hat{N}_2^{\alpha'}$ on detectors 1 and 2:
\begin{multline}
\label{eq:covariance}
\mathrm{cov}(\hat{N}^{\alpha}_1,\hat{N}^{\alpha'}_2)= \tau^2 \int\limits_0^\infty \!\! \int\limits_0^\infty \!\!\int\limits_0^\infty \!\!\int\limits_0^\infty\!\! \frac{d\omega}{\sqrt{2\pi}} \frac{d\omega'}{\sqrt{2\pi}} \frac{d\omega''}{\sqrt{2\pi}} \frac{d\omega'''}{\sqrt{2\pi}} \\  \left \langle  \hat{a}^{\alpha \dagger}(\omega) \hat{a}^{\alpha' \dagger}(\omega') \hat{a}^{\alpha'}({\omega''}) \hat{a}^{\alpha}({\omega'''}) \right \rangle \\ t_1^{*}(\omega) t_2^{*}(\omega') t_2(\omega'') t_1(\omega''') - \big\langle \hat{N_1}^\alpha \big\rangle \big\langle \hat{N_2}^{\alpha'}\big\rangle.
\end{multline}

In \eqref{eq:covariance} the expectation value is taken with respect to the \textit{in} vacuum state $\left|0^{\mathrm{in}}\right \rangle$ and $t_1$ and $t_2$ are the filters for detectors 1 and 2, respectively.
We calculate the fourth order moment of the field $\left \langle \hat{a}^{\alpha \dagger}(\omega) \hat{a}^{{\alpha'} \dagger}({\omega'}) \hat{a}^{\alpha''}({\omega''}) \hat{a}^{\alpha'''}({\omega'''}) \right \rangle$ in appendix \ref{app:corrfunctions}. 
Inserting the result \eqref{eq:4thmom2} into \eqref{eq:covariance} we find
\begin{widetext}
\begin{equation} 
\begin{split}
 \label{eq:covofS1}
 \mathrm{cov}&(\hat{N}^\alpha_1,\hat{N}^{\alpha'}_2) =\left(\frac{\tau}{2 \pi}\right)^2 \int\limits_0^\infty \int\limits_0^\infty  d\omega \, d\omega' \bigg[  \delta_{A} \sum_{\beta, \beta' \notin {\{\alpha\}} }  \mathcal{S}^*_{\alpha \beta'}(\omega) \mathcal{S}^*_{\alpha' \beta'}(\omega) \mathcal{S}_{\alpha' \beta}(\omega') \mathcal{S}_{\alpha \beta}(\omega')t_1^{}(\omega) t_2^{}(\omega) t_2(\omega') t_1(\omega') \\&+\delta_{S} \sum_{\beta, \beta' \notin \{{\alpha}\} }  \mathcal{S}^*_{\alpha \beta'}(\omega) \mathcal{S}^*_{\alpha' \beta}(\omega') \mathcal{S}_{\alpha' \beta'}(\omega) \mathcal{S}_{\alpha \beta}(\omega')t_1^{}(\omega) t_2^{}(\omega') t_2(\omega) t_1(\omega')  \\&+ \sum_{\beta\notin \{{\alpha}\}}\sum_{\beta'\notin \{{\alpha'}\}}  |S_{\alpha \beta}(\omega)|^2 |S_{\alpha' \beta'}(\omega')|^2 |t_1(\omega)|^2 |t_2(\omega')|^2 \bigg]-\big\langle \hat{N_1}^\alpha \big\rangle \big\langle \hat{N_2}^{\alpha'} \big\rangle .
\end{split}
\end{equation}
\end{widetext}
Here $\mathcal{S}_{\alpha \beta}$ is identical (complex conjugate) to $S_{\alpha \beta}$ if mode $\alpha$ is of positive (negative) norm. $\delta_S, \, (\delta_A)$ equals unity if the correlated modes $\alpha$ and $\alpha'$ have identical (opposite) norm and zero otherwise. Reverting from $\mathcal{S}$ back to $S$, we observe that in the $\delta_A$-term two $\mathcal{S}$ elements have to be conjugated and in the $\delta_S$-term either all or none. Since the result, as well as $t_i(\omega)$, are real, the first two sums in \eqref{eq:covofS1} are equal. Hence we observe that the mutually exclusive and exhaustive $\delta_S$ and $\delta_A$ let us combine the first two terms into one and the last two terms cancel due to \eqref{eq:flux} and \eqref{eq:hatN}:
\begin{equation} \label{eq:covofS} \mathrm{cov}(\hat{N}^\alpha_1,\hat{N}^{\alpha'}_2) =\left(\frac{\tau}{2 \pi}\right)^2  \bigg| \sum_{\beta \notin {\{\alpha\}} } \int\limits_\Delta d\omega \,{S}^*_{\alpha \beta}(\omega) {S}_{\alpha' \beta}(\omega) \bigg| ^2.
\end{equation}
$\Delta$ is the spectral overlap of the two detectors. This result allows us to quantify the spectrally resolved photon number correlations of any stationary process in quantum optics. The correlations are contained in the scattering matrix and are dependent on the spectral overlap of the detectors regarding the investigated mode. 

We also calculate the variance $\mathrm{var}(\hat{N}^\alpha_1)= \langle \hat{N}^\alpha_1 \hat{N}^\alpha_1 \rangle - \langle \hat{N}^\alpha_1 \rangle^2  $. Using eq. \eqref{eq:nnovariance} in an expression corresponding to \eqref{eq:covariance}, but not normally ordered and with $\alpha=\alpha'$, we obtain
\begin{equation}
\label{eq:variance}
\mathrm{var}(\hat{N}^\alpha_1)=\langle \hat{N}^\alpha_1 \rangle(\langle \hat{N}^\alpha_1 \rangle+\frac{\tau \Delta_1}{2 \pi}).
\end{equation}

Now that we have calculated the photon numbers and variances detected in certain modes, we briefly consider detecting photons from more than a single mode. Assuming the detector is sensitive for a frequency $\Omega$, then we can read off the dispersion diagram in Fig.\ref{fig:labdisprel}, that there are two mode solutions with $+\Omega$ and two solutions with $-\Omega$. Out of the four solutions exactly one has positive group velocity and positive moving frame frequency $\omega$. Thus the detector frequency can be identified with a unique mode in general \footnote{Exceptions are possible: (a) The frequency $\Omega$ separating two modes could lie inside the detected frequency interval. In this case the interval is reduced to the mode with group velocity away from the RIF. (b) Two out-LMs from either side of the RIF might share a laboratory frequency $\Omega$. In this case they typically do not share the moving frame frequency $\omega$ and the covariances and variances of the modes add up.}.
 
The photon-flux Pearson correlation coefficient between detectors 1 and 2, corresponding to modes $\alpha$ and $\alpha'$, is
\begin{widetext}
\begin{multline}
\label{eq:corrcoefficient}
\textrm{C}( \hat{N}^{\alpha}_1, \hat{N}^{\alpha'}_2)
=\frac{\mathrm{cov}(\hat{N}^\alpha_1,\hat{N}^{\alpha'}_2)}{[\mathrm{var}(\hat{N}^\alpha_1)\,\mathrm{var}(\hat{N}_2^{\alpha'})]^{1/2}}\\
= \frac{  \bigg|\sum\limits_{\beta \notin {\{\alpha\}} } \int\limits_\Delta \!d\omega \,{S}^*_{\alpha \beta} {S}_{\alpha' \beta} \bigg| ^2}{\bigg[ ( \sum\limits_{\beta \notin {\{\alpha\}} }\int\limits_{\Delta_1} \!\!d\omega |{S}_{\alpha \beta}|^2 )( \sum\limits_{\beta \notin {\{\alpha\}} }\int\limits_{\Delta_1} \!\!d\omega |{S}_{\alpha \beta}|^2+\Delta_1) ( \sum\limits_{\beta \notin {\{\alpha'\}} }\int\limits_{\Delta_2} \!\!d\omega |{S}_{\alpha' \beta}|^2) ( \sum\limits_{\beta \notin {\{\alpha'\}} }\int\limits_{\Delta_2} \!\!d\omega |{S}_{\alpha' \beta}|^2+\Delta_2) \bigg] ^{1/2}}\\
= \frac{\Delta^2}{\Delta_1 \Delta_2}\frac{  \bigg|\sum\limits_{\beta \notin {\{\alpha\}} }  \,{S}^*_{\alpha \beta} {S}_{\alpha' \beta} \bigg| ^2}{\bigg[ \sum\limits_{\beta \notin {\{\alpha\}} } |{S}_{\alpha \beta}|^2 ( \sum\limits_{\beta \notin {\{\alpha\}} } |{S}_{\alpha \beta}|^2+1)  \sum\limits_{\beta \notin {\{\alpha'\}} } |{S}_{\alpha' \beta}|^2 ( \sum\limits_{\beta \notin {\{\alpha'\}} } |{S}_{\alpha' \beta}|^2+1) \bigg] ^{1/2}}.
\end{multline}
\end{widetext}

In the last step we have assumed the scattering matrix to change little across the narrow detection bandwidth. The normalisation ensures that $|\textrm{C}( \hat{N}^{\alpha}_1, \hat{N}^{\alpha'}_2)|\leq1$. The correlations generated by the scattering are entirely positive, indicating their origin of entangled photon pair generation. From \eqref{eq:corrcoefficient} we can easily extract the (narrow bandwidth) self-correlation with matched filters
\begin{equation}
\label{eq:selfcorr}
\textrm{C}( \hat{N}^{\alpha}_1, \hat{N}^{\alpha}_1)
=\frac{\langle \hat{N}^{\alpha}_1 \rangle^2}{\mathrm{var}(\hat{N}^{\alpha}_1)}=1-\frac{\langle \hat{N}^{\alpha}_1 \rangle \frac{\tau \Delta}{2 \pi}}{\mathrm{var}(\hat{N}^{\alpha}_1)},
\end{equation}

which is a measure of the noise $\mathrm{var}(\hat{N}^{\alpha}_1)$ in mode $\alpha$  relative to the Poisson noise variance of  $\langle \hat{N}^{\alpha}_1 \rangle \frac{\tau \Delta}{2 \pi}$. Using eq. \eqref{eq:variance}, we see that $\textrm{C}$ increases from 0 (vacuum) and approaches 1 (maximum noise) for increasing photon numbers.

Furthermore, we calculate the second order correlation function $g^{(2)}_{\alpha \alpha'} = \frac{\langle:\hat{N}^{\alpha}_1 \hat{N}^{\alpha'}_2  : \rangle}{\langle \hat{N}^{\alpha}_1 \rangle\langle \hat{N}^{\alpha'}_2 \rangle}$. Again assuming narrowband detection, we obtain:
\begin{equation}
\label{eq:g2aa'}
	g^{(2)}_{\alpha \alpha'} = \frac{\Delta^2}{\Delta_1 \Delta_2} \left[1+ \frac{ |\sum\limits_{\beta \notin {\{\alpha\}} }  \,{S}^*_{\alpha \beta} {S}_{\alpha' \beta} |^2}{\sum\limits_{\beta \notin {\{\alpha\}} } |{S}_{\alpha \beta}|^2\sum\limits_{\beta \notin {\{\alpha'\}} } |{S}_{\alpha' \beta}|^2}\right]
\end{equation}
as well as the single detector correlation
\begin{equation}
\label{eq:g2aa}
	g^{(2)}_{\alpha \alpha} = 2.
\end{equation}
Here we recover the expected result that the scattering induces correlated noise between the two detectors, which both detect the statistics of chaotic light.

The photon numbers and correlations are calculated in the \textit{moving frame}. Photon numbers and time-bandwidth products $\tau \Delta$ are frame invariant. Therefore, re-interpreting $\tau$ and $\Delta$ as the \textit{laboratory} interaction time and detection bandwidth, the \textit{laboratory} frame correlation coefficient of modes $\alpha$  and $\alpha'$ is identical to eq. \eqref{eq:corrcoefficient}. All other photon number relations \eqref{eq:hatN}-\eqref{eq:g2aa} can be interpreted for the laboratory frame as well.

The \textit{moving frame} flux density $\phi^{\alpha}({\omega})$ in \textit{out} GM $\alpha$ was given in \eqref{eq:flux}. 
The \textit{laboratory frame} flux density $\Phi^{\alpha}(\Omega)$ is obtained from this by \cite{finazzi_quantum_2013}
\begin{equation}
\label{eq:LFrate}
\Phi^{\alpha}({\Omega})=\left(1-\frac{u}{v_{g}(\Omega)}\right)\phi^{\alpha}({\omega}).
\end{equation}
The total spectral density at $\Omega$ (or $\lambda$) $\Phi$ ($\Phi_\lambda$) is then found by adding contributions of all GMs  \cite{jacquet_quantum_2015,Jacquet_PhDThesis_2017} :
\begin{equation}
\begin{split}
\label{eq:LSD}
\Phi({\Omega})&=\sum_\alpha \Phi^{\alpha}({\Omega}) \\
\Phi_\lambda({\lambda})&= \frac{2\pi c}{\lambda^2} \sum_\alpha \Phi^{\alpha}({\frac{2\pi c}{\lambda}}).
\end{split}
\end{equation}
All other observables of the output field can be calculated from the scattering matrix. 

\section{Analytical calculation of the scattering}\label{sec:analyticsS}

We now derive the scattering matrix analytically.
 We continue to use the step in the index, as in Fig.\ref{fig:RIF}.
For a monochromatic field of frequency $\omega$ the conjugate momenta $\Pi_A$ and $\Pi_{P_i}$ as well as their first spatial derivative can be expressed by the electromagnetic field $A$, the polarisation fields $P_i$ as well as their derivatives in a homogeneous region using Eqs.\eqref{eq:fieldmotioneqMFFT1}, \eqref{eq:fieldmotioneqMFFT2} and \eqref{eq:fieldmotioneqMFFT4}. Therefore, we can express all fields by components of a vector $\vec{W}= (A, P_1, P_2, P_3, A', P'_1, P'_2, P'_3)^T$, which is continuous across the boundary. $A'$ etc. denotes the spatial derivative of $A$.

Similar to \eqref{eq:fieldtransfo} we create a matrix of the eight LM solutions $\vec{W}$ of the dispersion relation, which we combine in a matrix $W$ as
\begin{equation}
\label{eq:wmat}
W_{L/R}=\left(\vec{W}_{L/R}^{\alpha_1}\ \vec{W}_{L/R}^{\alpha_2}\ ... \ \vec{W}_{L/R}^{\alpha_8}\right),
\end{equation}
with $\vec{W}$'s ordered by laboratory-frame frequency (\textit{i.e.} modes $u,\; uo,\; mo,\;lo,\;l,\;nl,\;nol,\; nu$ in the case of 8 propagating modes).
$\vec{V}$ and $\vec{W}$ are related by \eqref{eq:fieldmotioneqMFFT1}-\eqref{eq:fieldmotioneqMFFT4} as:
\begin{widetext}
\begin{equation}
\label{eq:Umatrix}
\vec{V}=
\left(\begin{array}{cccccccc}
1&0&0&0&0&0&0&0 \\
0&1&0&0&0&0&0&0 \\
0&0&1&0&0&0&0&0 \\
0&0&0&1&0&0&0&0 \\
i\frac{\omega}{4\pi c^2}&0&0&0&0&0&0&0\\
\frac{\gamma}{c}&-i\frac{\omega\gamma}{\kappa_1\Omega_1^2}&0&0&0&-\frac{u\gamma^2}{\kappa_1\Omega_1^2}&0&0\\
\frac{\gamma}{c}&0&-i\frac{\omega\gamma}{\kappa_2\Omega_2^2}&0&0&0&-\frac{u\gamma^2}{\kappa_2\Omega_2^2}&0\\
\frac{\gamma}{c}&0&0&-i\frac{\omega\gamma}{\kappa_3\Omega_3^2}&0&0&0&-\frac{u\gamma^2}{\kappa_3\Omega_3^2}\\
\end{array}\right)
\vec{W},
\end{equation}
\end{widetext}
for a field at frequency $\omega$.
We call the matrix in \eqref{eq:Umatrix} $\mathcal{U}$, and note that $Det(\mathcal{U})=0$.
 We make \eqref{eq:Umatrix} applicable for global modes by defining $\mathcal{U}=\mathcal{U}_L\,\theta(-x)+\mathcal{U}_R\,\theta(x)$. Hence with \eqref{eq:fieldtransfo}, the scattering matrix is determined as  
\begin{eqnarray}
\label{eq:Wtransfo}
W^\mathrm{in}(x)&=&W^\mathrm{out}(x) \,S.
\end{eqnarray}

Complete sets of global \textit{in} and \textit{out} modes can now be written as:
\begin{eqnarray}
\label{eq:GMsets1}
W^\mathrm{in}(x)&=&W_L(x)\, \sigma_L^\mathrm{in} \,\theta(-x)+W_R(x)\, \sigma_R^\mathrm{in} \,\theta(x)\\
\label{eq:GMsets2}
W^\mathrm{out}(x)&=&W_L(x)\, \sigma_L^\mathrm{out} \,\theta(-x)+W_R(x)\, \sigma_R^\mathrm{out} \,\theta(x),
\end{eqnarray} 
where the $8\times8$ $\sigma$-matrices contain the coefficients of the 8 local modes for the 8 global modes. Insertion of \eqref{eq:GMsets1} and \eqref{eq:GMsets2} into \eqref{eq:Wtransfo} yields the scattering matrix 
\begin{equation}
\label{eq:smatrixsigma}
S={\sigma^{out}_{L}}^{-1}\sigma^{in}_{L}={\sigma^{out}_{R}}^{-1}\sigma^{in}_{R}.
\end{equation}
We arrive at an expression for the scattering matrix in terms of the mode coefficient matrices $\sigma$. The latter are calculated using the matching conditions.

The mode decompositions of global modes into local modes on the left and on the right of the interface, respectively, are related by the matching conditions at $x=0$. First we define $\vec{\bar{W}}$ similar to $\vec{\bar{V}}$ in  \eqref{eq:PWsolsMF}. For \textit{in} and \textit{out} modes we write
\begin{align}
\label{eq:matchINLeftRightSigma}
W^\mathrm{in}\Big| _{\substack{x=0 \\ \tau=0}}&=\bar{W}_{L}\ \sigma_{L}^{in}=\bar{W}_{R}\ \sigma_{R}^{in} \\
\label{eq:matchOUTLeftRightSigma}
W^\mathrm{out}\Big| _{\substack{x=0 \\ \tau=0}}&=\bar{W}_{L}\ \sigma_{L}^{out}=\bar{W}_{R}\ \sigma_{R}^{out}.
\end{align}
In Eqs.\eqref{eq:matchINLeftRightSigma} and \eqref{eq:matchOUTLeftRightSigma}, the $\sigma$'s are four $8\times8$ matrices. The way we construct the global modes defines half of the coefficients in $\sigma_L$ and $\sigma_R$ for \textit{in} and \textit{out} modes \footnote{For the unphysical (exponentially growing) mode, the construction is different: it is composed from the unphysical mode \textit{only} on one side. The associated GM serves as \textit{in} and identically as \textit{out} mode. Hence unphysical GMs scatter into themselves.}:
\begin{itemize}
\item when constructing an \textit{in} GM $\alpha$, we set to zero the coefficients of all remaining LMs, that propagate toward the RIF;
\item we contemplate wavepacket normalisation; considering the defining \textit{in} (\textit{out}) mode as a long wavepacket, it asymptotically for $\tau \rightarrow -\infty$ ($\tau \rightarrow \infty$) must have unit coefficient and therefore does so at all times.
\end{itemize}
This leaves us with 64 unknowns in either \eqref{eq:matchINLeftRightSigma} or \eqref{eq:matchOUTLeftRightSigma}. The $\bar{W}$'s are known, so we can obtain all $\sigma$'s from a single matrix
\begin{equation}
\label{eq:Amatrixdef}
A=\bar{W}_L^{-1}\ \bar{W}_R
\end{equation}
and calculate the scattering matrix by \eqref{eq:smatrixsigma}. Thus we have demonstrated how to completely characterize the scattering off a RIF analytically by calculation of the scattering matrix. 
In Appendix \ref{app:smfrommcs} we exemplify one mode configuration in detail (further calculations can be found in \cite{Jacquet_PhDThesis_2017}), and explicitly derive the \textit{in} and \textit{out} $\sigma$ matrices, and the corresponding scattering matrix.
In Appendix \ref{app:quasiunitarity}, we show that the Scattering Matrix is quasi-unitary, and thus correctly normalised.

We have now presented the analytical framework describing and characterizing the spontaneous emission of light at a moving index front in a dispersive medium.
Our method can be used with any single- or multi-pole dispersion relation that can be cast in the Sellmeier form \eqref{eq:SellmDispRelMF}.
Once the scattering matrix has been obtained, any observable can be computed.
In the final section of this paper, we compute realistic example spectra and correlation matrices, and identify characteristics of emission at the optical event horizon.

\section{Case study}\label{sec:numerics}

We now present a case study of a scattering matrix calculation. Optical analogue experiments are different from their fluid-based counterparts (such as BECs \cite{steinhauer_observation_2016} or water waves \cite{rousseaux_observation_2008,weinfurtner_measurement_2011,Rousseaux_PRL_2016}) in that the reference frames are exchanged: the rest frame of the optical experiment corresponds to the frame of the moving fluid and vice versa. In both analogues the measurements are performed in the laboratory frame. Therefore, in our case study, we  present spectra and mode-correlation maps in both frames for a comparison with the fluid-based analogues and explain the particularities of optical analogues.

 We consider the example of light in bulk fused silica at a step-like RIF. The Sellmeier coefficients are: $\kappa_{1,2,3}= 0.07142$, $0.03246$, and $0.05540$ for the elastic constants, and $\Omega_{1,2,3}=190.341 \, \mathrm{THz}$, $16.2047\, \mathrm{PHz}$, and $27.537\, \mathrm{PHz}$ for the resonance frequencies \cite{Agrawal_2012}.
\begin{figure}[t]
\centering
\includegraphics[width=.95\columnwidth]{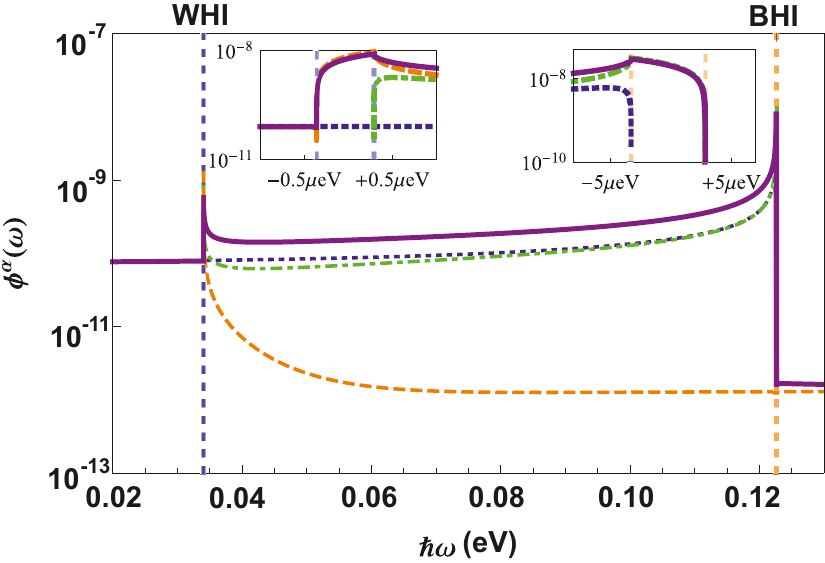}
\caption[Emission spectra of the four modes in the moving frame]{Emission spectra of the four optical modes in the moving frame for $\delta n=2\times10^{-6}$. The photon fluxes of emission: mode \textit{noL}, purple solid line; \textit{loL}, orange dashed line; \textit{uoL}, blue dotted line; \textit{moR}, green dot-dashed line. Insets show zoomed-in regions around the white- and black hole intervals.\label{fig:mfflux}}
\end{figure}
In \cite{jacquet_quantum_2015} we had calculated emission spectra for a number of very high step heights. Here we perform computations for a moderate step-height of $\delta n=2\times10^{-6}$ as in \cite{philbin_fiber-optical_2008,choudhary_efficient_2012,Jacquet_PhDThesis_2017}.

As can be seen in Fig.\ref{fig:mfflux}, the spontaneous emission from the RIF is peaked in two very narrow frequency intervals.
The low-frequency interval corresponds to white hole emission ($\phi^{noL}({\omega})\sim 10^{-9}$ photons per time-bandwidth) and the large-frequency interval to black hole emission ($\phi^{noL}({\omega})\sim 10^{-8}$). In \cite{robertson_hawking_2011, jacquet_quantum_2015} it was shown that the spectrum has a characteristic `shark fin' shape, a sharp increase of emission as $\omega$ approaches the black hole interval and a slower decrease inside  that interval.
Despite the narrow width of the peaks, their `shark fin' shape remains, as seen in the insets of Fig.\ref{fig:mfflux}. This shape is a signature of horizon-like emission.
Over the narrow white hole interval (WHI) indicated in Fig.\ref{fig:mfflux}, emission is mainly into partnered modes \textit{noL} (purple line) and \textit{loL} (orange-dashed line). 
Over the black-hole interval (BHI), emission is strongest into modes \textit{noL} and \textit{moR} (green-dot-dashed line). The photon pairs produced have partners of opposite norm and the horizon emission follows the kinematics explored in Fig.\ref{fig:intervals}. Because \textit{noL} is the only negative norm mode of optical frequency, the flux into this mode is strong. Note that the non-optical mode \textit{nlL} is contributing weakly at the WHI.

This can be assessed further by computing the matrix of photon number correlations between all modes, including non-optical modes, by eq. \eqref{eq:corrcoefficient}. 
\begin{figure}[t]
\centering
\includegraphics[width=.95\columnwidth]{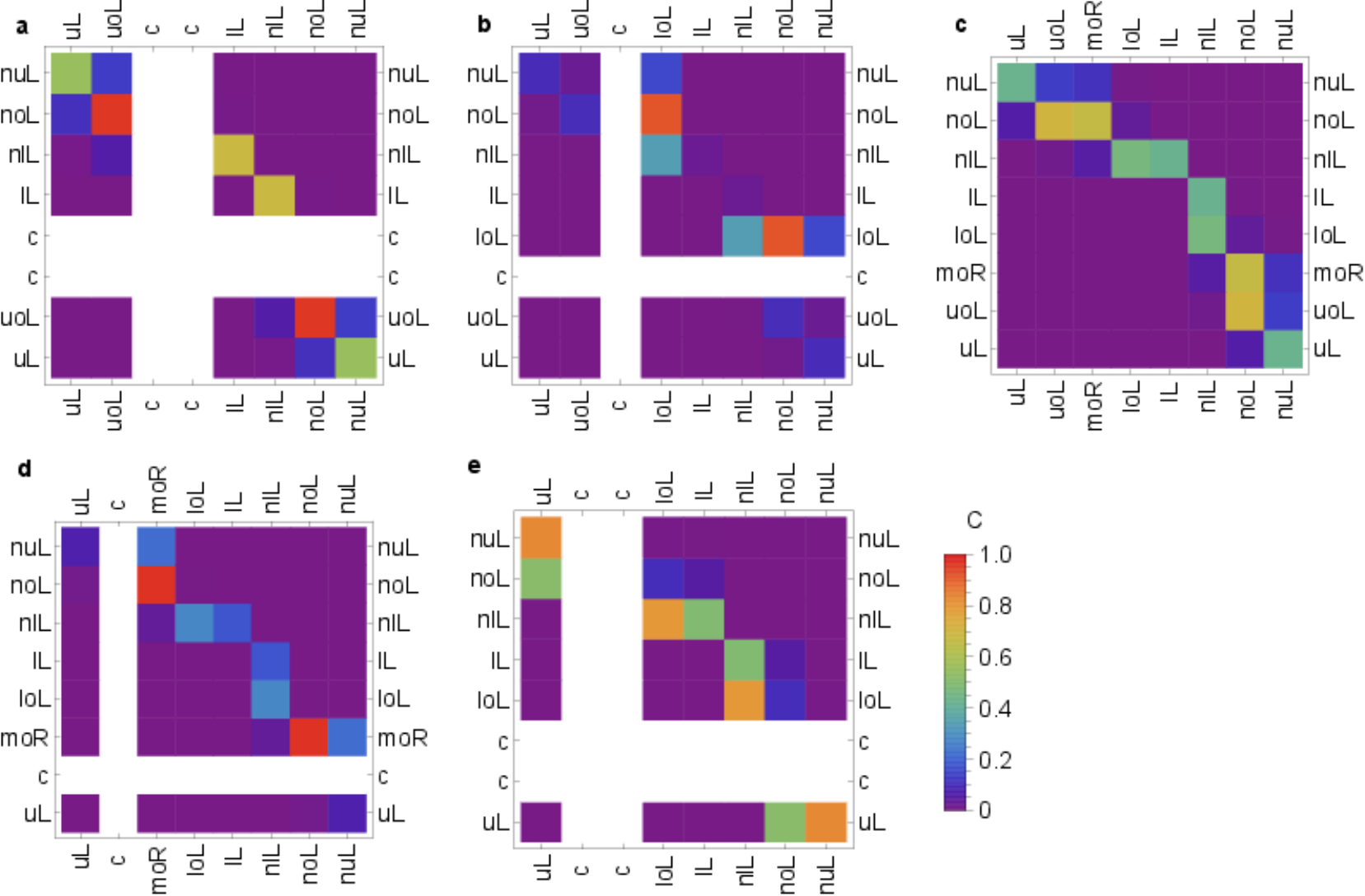}
\caption[Mode-solutions cross correlations]{Photon number correlations according to modes. Correlations between the 8 \textit{out} GMs in the five frequency intervals of Fig.\ref{fig:intervals}: a) $\omega=0.027 eV/\hbar\leq\omega_{minL}$; b) white hole interval, $\omega=(\omega_{minL}+\omega_{minR})/2$ ; c) $\omega=(\omega_{minR}+\omega_{maxL})/2$; d) black hole interval, $\omega=(\omega_{maxL}+\omega_{maxR})/2$  ; e) $\omega=0.147 eV/\hbar>\omega_{maxR}$. $\delta n=2\times10^{-6}$ throughout. No correlations are shown for unphysical complex modes. \label{fig:mfcorr}}
\end{figure}
In  Fig.\ref{fig:mfcorr} we observe correlations between negative and positive norm modes for typifying frequencies $\omega$.
Correlations are generally strongest between the optical modes. Because \textit{noL} is the unique negative-norm optical mode, we find strong correlations between this mode and positive-norm optical modes. The correlations, which are independent of the fluxes in the modes, are different if horizons exist (Fig.\ref{fig:mfcorr} (b), (d)).  Over the white- and black hole intervals, there is a single large photon number correlation between \textit{noL} and \textit{loL}, and \textit{noL} and \textit{moR}, respectively, with other correlations being small or zero. The pairs of modes correspond to the Hawking emission and the partner as expected  from the WHI and the BHI in Fig.\ref{fig:intervals}.
Without horizons (Fig.\ref{fig:mfcorr} (a), (c), (e)), strong correlations exist between typically three mode pairs simultaneously. These involve non-optical modes, although the emission is weak. 
Note that also extremely weak positive-positive norm correlations exist which are not visible in the figure, e.g. between \textit{uoL} and \textit{moR}. 

To summarise,  the flux of spontaneous emission is dominated by white- or black hole-horizon physics and drops significantly beyond that.
Hence, the spectral characteristic is a `shark fin' shape.
Over the analogue white- and black hole intervals,  pair-wise emission at optical frequencies dominates.
Clearly, these are signature effects of horizon physics in dispersive (optical) media.

\begin{figure}[t]
\centering
\includegraphics[width=\columnwidth]{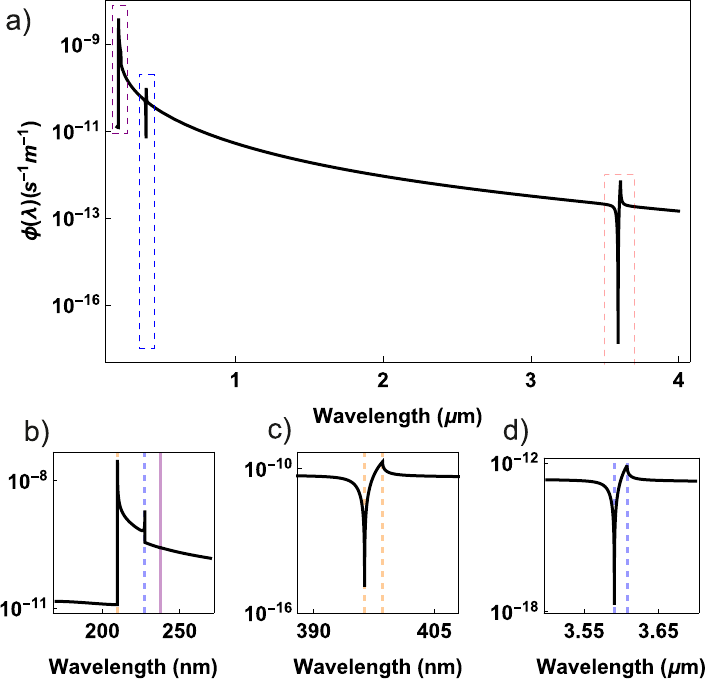}
\caption[Spectral density of  emission in the laboratory frame]{Spectral density of  emission in the laboratory frame. At each wavelength the total photon flux density is the sum of contributions from all modes (Eq.\eqref{eq:LSD}). The spectrum is computed for $\delta n=2\times10^{-6}$ and  $u=2/3\,c$ in bulk fused silica. Emission from horizons is found at these dashed lines: $210\,$nm: black hole (orange), $227\,$nm: white hole (blue), $398\,$nm: black hole partner (orange), $3.6\,\mu$m: white hole partner (blue). There is a phase velocity horizon at $\lambda_{pvh}=237$nm (purple).\label{fig:LSD}}
\end{figure}

We compute now these signature effects as they would be observed in the laboratory frame.
We begin with the spectral density \eqref{eq:LSD}  shown in Fig.\ref{fig:LSD} for $\delta n=2\times10^{-6}$.
Similar to its moving-frame counterpart, the spectrum consists of intervals of white hole, black hole and horizonless emission.
Emission is peaked over analogue white- and black-hole intervals. The peak at $210\,$nm corresponds to black hole emission into mode \textit{noL}.
The peak at $227\,$nm is white hole emission into the same mode. For longer wavelengths this spectral density decreases until $\lambda_{vm}=396\,$nm.
At this wavelength, the black hole intervals for \textit{uoL} and \textit{moR} (at short and long wavelength, respectively) overlap slightly.
The laboratory frame spectral density dips down about 5 orders of magnitude at $\lambda_{vm}$. This feature corresponds to the sharp black hole peak (the fall after the peak) in Fig.\ref{fig:mfflux} \footnote{At the edges of the horizon-like intervals the group-velocity of the \textit{out} GM vanishes in the moving frame and emission drops to zero.}.
Finally, we observe peaks at $398\,$nm and around $3.6\,\mu m$ that account for the black hole and white hole emission into positive modes \textit{moR} and \textit{loL}, respectively.
All peaks exhibit the signature `shark fin' structure of horizon emission, as visible in the insets.

The various spectral peaks and dips we have identified have a narrow linewidth: about $1\,$nm below $250\,\mathrm{nm}$, $2\,$nm around  $400\,\mathrm{nm}$ and $17\,$nm in the IR.
They are associated with strong spectral correlations amongst distinct spectral intervals.
Measuring these photon-number correlations convincingly reveals their horizon physics origin. 

\begin{figure}[t]
\centering
\includegraphics[width=\columnwidth]{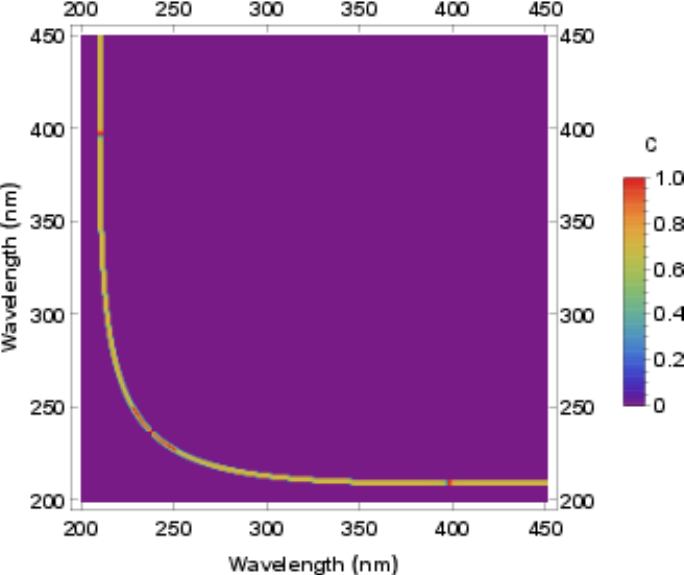}
\caption[Correlation map]{Photon number correlations in the laboratory: A refractive index step ($\delta n=2\times10^{-6}$) is moving at $u=2/3\,c$ in bulk fused silica, leading to spontaneous emission of entangled photon pairs.\label{fig:lfcorr}}
\end{figure}

Fig.\ref{fig:lfcorr} presents the photon number-correlation coefficient \eqref{eq:corrcoefficient} across the spectrum of laboratory frame emission.
Coefficients between different modes display a single continuous contour of significant photon number correlations across the entire spectrum. 
The contour indicates correlations between the negative-norm mode \textit{noL} (below $237\,\mathrm{nm}$) and the positive-norm modes \textit{uoL} (between $237\,\mathrm{nm}$ and $396\,\mathrm{nm}$) or \textit{moR} (beyond $396\,\mathrm{nm}$). The contour thus indicates where these mode pairs share a common moving frame frequency $\omega$. Note that this depends on dispersion in a nontrivial way and is not a hyperbolic relation. 
The correlation coefficients of Fig.\ref{fig:mfcorr} and Fig.\ref{fig:lfcorr} are mostly identical for identical frequencies $\omega$. Along the contour, the  correlation coefficients gradually decrease as the wavelength of \textit{noL} decreases and the wavelength of \textit{uoL}, and subsequently of \textit{moR}, increases.
We observe the strongest correlations between  $210\, \textrm{nm}$ and $397\, \textrm{nm}$, corresponding to pair-wise emission into \textit{noL} and \textit{moR} at the black hole horizon.
Here, the correlation coefficient is 0.97, indicating that the quantum state at the output is almost a pure two-mode squeezed vacuum.
It deviates from unity because of weak correlations of mode \textit{moR} with non-optical modes (Fig.\ref{fig:mfcorr}). On the diagonal, the self-coefficients \eqref{eq:selfcorr} characterize the photon number noise relative to Poisson noise. These elements are very small as individual modes carry chaotic noise (cf. \eqref{eq:g2aa}) and the photon numbers are small. In a two-mode squeezed vacuum, the constituent modes are in a thermal state. In addition, the effective temperature of the distribution depends on frequency due to dispersion. 

Pair-wise emission also dominates over the white-hole interval.
Correlated pairs of photons are emitted into modes \textit{noL} and \textit{loL}, respectively. The correlation coefficient is 0.92 only, because some photons in mode \textit{loL} are paired with partners in a non-optical mode. The negative norm partner photon can be found in any one of the three negative norm modes. Mode \textit{loL} is situated at $3.6\,\mathrm{\mu m}$ and mode \textit{noL} at $227\,\mathrm{nm}$.
The very strong correlation between outgoing modes separated by a horizon is characteristic of the Hawking effect.


\begin{table*}\setlength{\arrayrulewidth}{0.3mm}
\begin{center}
\begin{minipage}[t]{130mm}\scriptsize
\begin{tabular}[b]{ |c|c|c|c|c|c|c|c|c|c|c|c| }
\hline \hline
\multicolumn{2}{|c|}{RIF}    &   \multicolumn{5}{c|}{White hole}   & \multicolumn{5}{c|}{Black hole}  \tabularnewline
 $\lambda_c$ (nm) & $u/c$  &   $\lambda_{noL}$ (nm) & $\lambda_{loL}$ ($\mu$m) & $\Phi^{noL}(\lambda)$ & $\Phi^{loL}(\lambda)$ &C & $\lambda_{noL}$ (nm) & $\lambda_{moR}$ (nm) & $\Phi^{noL}(\lambda)$ & $\Phi^{moR}(\lambda)$ & C\tabularnewline  \hline
 $400$ & $2/3$ & $227$ & $3.6$ & $3.5\times10^{13}$ & $7.8\times10^8$ & $0.92$ & $209.8$ & $398.5$ & $2.1\times10^{14}$ & 
 $1.9\times10^{11}$ & $0.97$\tabularnewline
 $800$ & $2.04/3$ & $379$ & $2.01$ & $3.4\times10^{13}$ & $1.5\times10^{10}$ & $0.99$ & $372$ & $810$ & $1.1\times10^{14}$ & $1.3\times10^{11}$ & $0.99$\tabularnewline
$1260$ & $2.05/3$ & $438$ & $1.36$ & $2.4\times10^{14}$ & $6.1\times10^{10}$ & $0.99$ & $438$ & $1317$ & $2.1\times10^{14}$ & $1.9\times10^{10}$ & $0.99$\tabularnewline
 $1990$ & $2.04/3$ & $379$ & $2.01$ & $6.7\times10^{13} $ & $1.5\times10^{10}$ & $0.99$ & $372$ & $810$ & $1.2\times10^{14}$ & $1.4\times10^{11}$ & $0.99$\tabularnewline 
 \hline
\end{tabular}
\end{minipage}
\caption{Dependence of horizon physics on the velocity $u$ of the RIF. $\lambda_c$: effective RIF central wavelength.  \label{tab:tableu}}
\end{center}
\end{table*}


For the low index step used in this paper, the $\omega$-intervals of black- and white hole horizons are very narrow and close to the two frequencies that have the same group velocity $u$ as the RIF. The RIF might be generated by an optical pulse via the optical Kerr effect. The velocity $u$ then corresponds to the group velocity of that pulse. Changing $u$, e.g. by shifting the pulse frequency, will therefore alter the wavelengths of the black- and white hole emission.
In Table \ref{tab:tableu} we show this influence for four pulse wavelengths, $\lambda_c=400$, $800$, $1260$ and $1990\,$nm. While $1260\,$nm is the zero dispersion wavelength, where the group index assumes a maximum, $800\,$nm and $1990\,$nm share the same group velocity. The table shows that overall emission frequencies for the optical modes are shifted, but the emission amplitudes or correlation strengths are largely robust.

Because the black hole emission into mode \textit{moR} is very close to the pulse wavelength, it will be difficult to distinguish it from photons of the pulse. Also, the white hole emission at near-infrared wavelengths is hard to detect. 
These problems can be overcome by using a similar medium, but with a shifted zero-dispersion wavelength, as well as a finite-length pulse instead of a step. In this case, the emission close to the group velocity-matched wavelength might lie in the visible/UV range. The emission from the front (back) of the pulse will be found in this range where the pulse edge constitutes a black (white) hole. At the same time the pulse can propagate in the anomalous group velocity regime where temporal self-focussing aids the maintenance or creation of a steep pulse edge \cite{Agrawal_2012}. The shift of the zero-dispersion wavelength can be achieved by use of metamaterial waveguides, such as photonic crystal fibres \cite{russell_photonic_2003,philbin_fiber-optical_2008,choudhary_efficient_2012,Jacquet_PhDThesis_2017}.  

\section{Discussion and conclusion}\label{sec:conclu}


We considered the mixing of modes of positive- and negative-norm at a horizon in dispersive, optical media.
We presented an analytical method for calculating the properties of spontaneous emission resulting from the mode-mixing process based on  \cite{finazzi_quantum_2013} and the Hopfield model \cite{hopfield_theory_1958,fano_atomic_1956,schutzhold_dielectricbh_2002}. Hence we derived the scattering matrix, that describes the mixing of modes, at a moving, step-like  RIF. The RIF may act as a black- or white hole horizon or as a horizonless emitter.
We quantify the emission with the spectral flux and the calculated photon number correlations.

We used the analytical calculation of the scattering matrix to compute, in a case study, spectra and correlations of emission.
We obtained a broad and  structured spectrum with emission peaks. The emission is in entangled photon pairs and occurs strongly from horizons on a background of a broadband, weak and horizonless emission. From these findings emerges the role of the horizon as an increase in the emission accompanied with an increase in photon number correlations, that are approaching unity.
The spectra and correlations are of principal importance for an experimental implementation. The resulting spectral peak, spectral correlations and their dependence on RIF velocity and medium dispersion are key identifiers of the types of spontaneous emission at the horizon by the Hawking effect or the horizonless emission.

The method can be generalized to study other RIF geometries, such as a finite-length pulse, and to calculate the spectra and spectral correlations. This is essential to identify the optimal conditions to observe the analogue Hawking effect. Importantly, the method can be used to provide a theory trace for optical analogue experiments and to study a variety of effects such as cosmological pair creation, analogue wormholes, and others \cite{carusotto_quantum_2013}.

\begin{acknowledgments}
The authors are thankful to Bill Unruh, Joseph Cousins, Stephen Barnett, Vyome Singh and Maya Petkova for insightful conversations. This work was funded by the EPSRC via Grant No. EP/L505079/1.
\end{acknowledgments}

\appendix
\section{Matching conditions of the fields at the RIF \label{app:matchingconditions}}

Here we derive the continuity of the fields across the RIF. On physical grounds, we consider the field, polarisation field, all conjugate momenta and their time derivatives to be finite.
By construction of the model \cite{schubert_nonlinear_1986}, the elastic constant $\kappa$ is discontinuous  and the inertia of the polarisation fields $(\kappa \Omega^2)^{-1}$ is continuous at the interface between the two homogeneous regions (cf.   \eqref{eq:nlindex}). For fields of single frequency  $\omega$ the field's time derivatives do not alter the spatial continuity of the fields. 
In the near-interface region, we integrate the equations of motion over space.
We begin with \eqref{eq:fieldmotioneqMFFT3}: 
\begin{equation}
\label{eq:spaceintegralconjugatemomentumemfield}
\begin{split}
\int_{-\epsilon_1}^{+\epsilon_2}i\omega\,\Pi_A\mathrm{dx}&=\int_{-\epsilon_1}^{+\epsilon_2}\frac{A''}{4\pi}\mathrm{dx}\\
&\ \ \ +\int_{-\epsilon_1}^{+\epsilon_2}\sum_{i=1}^3\frac{\kappa_i\Omega^2}{\gamma c}\left(\Pi_{P_i}-\gamma\frac
{A}{c}\right)\mathrm{dx}.
\end{split}
\end{equation}
All finite terms integrate to zero in the limits $(\epsilon_1\rightarrow0, \epsilon_2\rightarrow0)$ and so 
\begin{equation}
\label{eq:emfieldcontinuity}
\lim_{\substack{\epsilon_1 \to 0 \\ \epsilon_2 \to 0}}\int_{-\epsilon_1}^{+\epsilon_2}\frac{A''}{4\pi}\,\mathrm{dx}=0.
\end{equation}
Thus $A''$ is finite and the vector potential $A$ is continuously differentiable at $x=0$: $A (0^-,t)=A (0^+,t)$ and $A' (0^-,t)=A' (0^+,t)$.
Proceeding similarly with  \eqref{eq:fieldmotioneqMFFT2} leads to 
\begin{equation}
\label{eq:polfieldcontinuity}
\begin{split}
\lim_{\substack{\epsilon_1 \to 0 \\ \epsilon_2 \to 0}}\int_{-\epsilon_1}^{+\epsilon_2}P'_i\,\mathrm{dx}=0 \quad
\Rightarrow \quad P_{i}(0^-,t)=P_{i}(0^+,t),
\end{split}
\end{equation}
i.e. the polarisation fields are continuous at $x=0$. Analogously we find from   \eqref{eq:fieldmotioneqMFFT4} that the conjugate momenta $\Pi_{P_i}$'s are continuous.
We established now the continuity of $A$, $P_i$, $\Pi_{P_i}$, and $\kappa_i \Omega^2_i$  in  \eqref{eq:fieldmotioneqMFFT2}, and so $P'_i$, too, is continuous: $P_i' (0^-,t)=P_i' (0^+,t)$.
Finally, turning back to  \eqref{eq:fieldmotioneqMFFT4}, in which $P_i$ and $\dot{\Pi}_{P_i}$  are continuous, we see that discontinuity of $\kappa_i$ implies discontinuity of  $\Pi'_{P_i}$.
Subtracting the rhs of  \eqref{eq:fieldmotioneqMFFT4} on either side of the RIF at $x = 0$ we obtain:
\begin{equation}
\label{eq:discontinuityofspatialconjugatepolfield}
 \Pi'_{P_i }(0^+,t)=\Pi'_{P_i }(0^-,t)+\frac{P_i}{u}\left(\frac{1}{\kappa_{i,R}}-\frac{1}{\kappa_{i,L}}\right).
\end{equation}

\section{Scattering matrix from the matching conditions\label{app:smfrommcs}}

In this appendix we explicitly derive the scattering matrix for the mode configuration of Fig.\ref{fig:intervals}(c). This corresponds to a frequency $\omega$ with 8 propagating modes on either side of the RIF interface.
We use \eqref{eq:matchINLeftRightSigma}  and \eqref{eq:Amatrixdef} to relate the $\sigma^{\mathrm{in}}$-matrices on either side. Each column of $\sigma^{\mathrm{in}}$ contains the coefficients of individual local modes to one of the global modes on that side of the RIF. We arrange the global and local modes, respectively, in decreasing order of $\Omega$, i.e. first \textit{$u^{\mathrm{in}}$}, then \textit{$uo^{\mathrm{in}}$} and so on.
We use the construction of GMs described above eq. \eqref{eq:Amatrixdef} and hence  \eqref{eq:matchINLeftRightSigma} writes:
\begin{equation}
\label{eq:sigmamatricesoscosc}
\renewcommand*{\arraystretch}{0.9}
\left(\begin{array}{cccccccc}
\multicolumn{8}{c}{\hfill}\\[0.0em]\multicolumn{8}{c}{\hfill}\\[0.0em]0&0&1&0&0&0&0&0\\[0.0em] \multicolumn{8}{c}{\hfill} \\[0.0em] \multicolumn{8}{c}{\hfill} \\[0.0em] \multicolumn{8}{c}{\hfill} \\[0.0em] \multicolumn{8}{c}{\hfill} \\[0.0em] \multicolumn{8}{c}{\hfill}  \end{array}\right)=
A
\left(\begin{array}{cccccccc}
1&0&0&0&0&0&0&0\\
0&1&0&0&0&0&0&0\\ \multicolumn{8}{c}{\hfill}
\\
0&0&0&1&0&0&0&0\\
0&0&0&0&1&0&0&0\\
0&0&0&0&0&1&0&0\\
0&0&0&0&0&0&1&0\\
0&0&0&0&0&0&0&1\end{array}\right)
\end{equation}
There are 64 unknowns, `empty' components of the matrices. We obtain:
\begin{widetext}
\begin{eqnarray}
\label{eq:sigmainL8x8}
\sigma_L^{in}&=&
\left(\begin{array}{ccccc}
A_{11}-\frac{A_{13}A_{31}}{A_{33}}&A_{12}-\frac{A_{13}A_{32}}{A_{33}}&\frac{A_{13}}{A_{33}}&A_{14}-\frac{A_{13}A_{34}}{A_{33}}&\cdots\\
A_{21}-\frac{A_{23}A_{31}}{A_{33}}&A_{22}-\frac{A_{23}A_{32}}{A_{33}}&\frac{A_{23}}{A_{33}}&A_{24}-\frac{A_{23}A_{34}}{A_{33}}&\cdots\\
0&0&1&0&\cdots\\
A_{41}-\frac{A_{43}A_{31}}{A_{33}}&A_{42}-\frac{A_{43}A_{32}}{A_{33}}&\frac{A_{43}}{A_{33}}&A_{44}-\frac{A_{43}A_{34}}{A_{33}}&\cdots\\
\vdots&\vdots&\vdots&\vdots&\ddots\\
A_{81}-\frac{A_{83}A_{31}}{A_{33}}&A_{82}-\frac{A_{83}A_{32}}{A_{33}}&\frac{A_{83}}{A_{33}}&A_{84}-\frac{A_{83}A_{34}}{A_{33}}&\cdots
\end{array}\right) \\ \vspace{10mm}
\label{eq:sigmainR8x8}
\sigma_R^{in}&=&
\left(\begin{array}{ccccc}
1&0&0&0&\cdots\\
0&1&0&0&\cdots\\
-\frac{A_{31}}{A_{33}}&-\frac{A_{32}}{A_{33}}&\frac{1}{A_{33}}&-\frac{A_{34}}{A_{33}}&\cdots\\
0&0&0&1&\cdots\\
\vdots&\vdots&\vdots&\vdots&\ddots\\
0&0&0&0&\cdots
\end{array}\right).
\end{eqnarray}
\end{widetext}

For the \textit{out} modes, \eqref{eq:matchINLeftRightSigma} is
\begin{equation}
\label{eq:sigmamatricesoscoscout}
\renewcommand*{\arraystretch}{0.9}
\left(\begin{array}{cccccccc}
1&0&0&0&0&0&0&0\\
0&1&0&0&0&0&0&0\\ \multicolumn{8}{c}{\hfill}
\\
0&0&0&1&0&0&0&0\\
0&0&0&0&1&0&0&0\\
0&0&0&0&0&1&0&0\\
0&0&0&0&0&0&1&0\\
0&0&0&0&0&0&0&1\end{array}\right)
=A\left(\begin{array}{cccccccc}
\multicolumn{8}{c}{\hfill}\\ \multicolumn{8}{c}{\hfill}\\0&0&1&0&0&0&0&0\\ \multicolumn{8}{c}{\hfill} \\  \multicolumn{8}{c}{\hfill}\\  \multicolumn{8}{c}{\hfill}\\  \multicolumn{8}{c}{\hfill}\\ \multicolumn{8}{c}{\hfill}\end{array}\right),
\end{equation}
and by comparison with \eqref{eq:sigmamatricesoscosc} we exchange L $\longleftrightarrow$ R and replace $A$ by $A^{-1}$ in \eqref{eq:sigmainL8x8} and \eqref{eq:sigmainR8x8}.  Furthermore, we invert $\sigma_L^{\mathrm{out}}$:
\begin{equation}
\label{eq:sigmaoutLTandTinv8x8}
{{\sigma_L^{out}}}^{-1}=\left(\begin{array}{ccccc}
1&0&0&0&\cdots\\
0&1&0&0&\cdots\\
A_{31}^{-1}&A_{32}^{-1}&A_{33}^{-1}&A_{34}^{-1}&\cdots\\
0&0&0&1&\cdots\\
\vdots&\vdots&\vdots&\vdots&\ddots\\
0&0&0&0&\cdots
\end{array}\right).
\end{equation}
Finally, by \eqref{eq:smatrixsigma}, we obtain the scattering matrix 
\begin{widetext}
\begin{equation}
\label{eq:Smatrix8x8}
\begin{split}
S=\left(\begin{array}{ccccc}
A_{11}-\frac{A_{13}A_{31}}{A_{33}}&A_{12}-\frac{A_{13}A_{32}}{A_{33}}&-\frac{A_{13}}{A_{33}}&A_{14}-\frac{A_{13}A_{34}}{A_{33}}&\cdots\\
A_{21}-\frac{A_{23}A_{31}}{A_{33}}&A_{22}-\frac{A_{23}A_{32}}{A_{33}}&-\frac{A_{23}}{A_{33}}&A_{24}-\frac{A_{23}A_{34}}{A_{33}}&\cdots\\
-\frac{A_{31}}{A_{33}}&-\frac{A_{32}}{A_{33}}&\frac{1}{A_{33}}&-\frac{A_{34}}{A_{33}}&\cdots\\
A_{41}-\frac{A_{43}A_{31}}{A_{33}}&A_{42}-\frac{A_{43}A_{32}}{A_{33}}&-\frac{A_{43}}{A_{33}}&A_{44}-\frac{A_{43}A_{34}}{A_{33}}&\cdots\\
\vdots&\vdots&\vdots&\vdots&\ddots\\
A_{81}-\frac{A_{83}A_{31}}{A_{33}}&A_{82}-\frac{A_{83}A_{32}}{A_{33}}&-\frac{A_{83}}{A_{33}}&A_{84}-\frac{A_{83}A_{34}}{A_{33}}&\cdots
\end{array}\right).
\end{split}
\end{equation}
\end{widetext}
In \eqref{eq:Smatrix8x8}, we have completed the derivation of the S matrix for mode configuration (c) with 8 propagating modes on either side of the interface. The A coefficients are taken from \eqref{eq:Amatrixdef}, i.e. from the normalized local mode components.
This derivation follows on from the matching conditions for the fields and their first spatial derivative at the interface and results in a straightforward expression that can easily be evaluated on a computer.

\section{Quasi-unitarity of the scattering matrix}\label{app:quasiunitarity}

The scattering matrix describes the basis change between  \textit{in}  and \textit{out} global modes. Both are orthonormal bases, at least when there are 8 propagating modes. In order to preserve orthonormality, the scattering matrix is constrained. Alternatively, this can be seen as the preservation of the commutator relation for \textit{in}  and \textit{out} annihilation and creation operators.

The orthonormality of the \textit{in} GMs $\alpha$ and $\alpha'$ is defined by a matrix $g$, 
\begin{equation}
\label{eq:gdef}
g_{\alpha\,\alpha'}=\left\langle \vec{\mathcal{V}}^{\mathrm{in}\,\alpha},\vec{\mathcal{V}}^{\mathrm{in}\,\alpha'}\right\rangle, 
\end{equation}
with respect to the scalar product \eqref{eq:scalarproduct}. The only non-zero elements  of $g$ are +1 or -1 on the diagonal, indicating the positivity or negativity of the mode norm. In the \textit{out} basis we have the same number of negative modes, because the norm is conserved during scattering, and so relation \eqref{eq:gdef} is also valid for \textit{out} GMs, if we order the modes accordingly. Using \eqref{eq:fieldtransfo} we calculate\begin{equation}
\label{eq:derivequasiunitarity}
\begin{split}
g&=\frac{i}{\hbar}\int dx\, \mathcal{V}^{\mathrm{in}\, \dagger} \,\left(\begin{array}{cc}0&I_4\\-I_4&0\end{array}\right)\, \mathcal{V}^{\mathrm{in}} \\
&=\frac{i}{\hbar}\int dx\, S^{\dagger}\mathcal{V}^{\mathrm{out}\, \dagger} \,\left(\begin{array}{cc}0&I_4\\-I_4&0\end{array}\right)\,  \mathcal{V}^{\mathrm{out}}S\\
&=S^{\dagger}g\,S.
\end{split}
\end{equation}
This relation is called `quasi-unitarity' and means that $S$ (and $S^{\dagger}$) is a member of the indefinite unitary group $U(5,3)$. We can reformulate this condition as a normalization condition for the rows (and columns) of $S$ as:
\begin{equation}
\label{eq:Snorm}
\sgn({\Omega^{\alpha}})=\sum_{\alpha' \in P}\left|S_{\alpha \alpha'}(\omega)\right|^2-\sum_{\alpha' \in N}\left|S_{\alpha {\alpha'}}(\omega)\right|^2,
\end{equation}
where $\sgn$ indicates the frequency sign and thus the norm of mode $\alpha$.
Ensuring that the scattering matrix is quasi-untiary is a useful test for numerical implementations.
The procedure is easily generalised for other cases, where there are complex, non-propagating mode solutions. As these generate non-physical modes, the scattering matrix becomes a block matrix and, although not normalizable, the norm of the unphysical mode can be defined as unity in $g$.

\section{Higher order correlation function}\label{app:corrfunctions}

In this appendix, we detail the calculation of photon number variances and covariances. These are expressed by the expectation value of the second and fourth order moments of the \textit{out} annihilation operators, which we calculate here. The expectation value is taken with respect to the \textit{in} vacuum state. Therefore, we write out the Bogoljubov transformation \eqref{eq:operatortransfo}, which connects \textit{in} and \textit{out} operators, in the norm-independent way:
\begin{equation}
	\label{eq:Bogoljubov}
	\hat{a}^{\mathrm{out}\,\alpha}(\omega)=\!\!\!\!\sum\limits_{\beta \in \{ \alpha \}}\!\!\mathcal{S}_{\alpha\beta}(\omega)\, \hat{a}^{\mathrm{in}\,\beta}(\omega)+\!\!\!\!\sum\limits_{\beta \notin \{ \alpha \}}\!\!\mathcal{S}_{\alpha\beta}(\omega
) \,\hat{a}^{\mathrm{in}\,\beta\, \dagger}(\omega) ,
\end{equation}
where $\{ \alpha \}$ again stands for the set of modes with norm identical to $\alpha$. The matrix $\mathcal{S}$ is equal to the scattering matrix $S$ except for the rows which belong to negative norm modes, that are complex conjugated. Creation operators for the \textit{out} modes are then obtained by Hermitian conjugation of \eqref{eq:Bogoljubov} only. Note that this expression is valid for any mode $\alpha$, whether of positive or negative norm.

We start with the second moment 
\begin{align}
\label{eq:2ndmom1}
		&\langle 0^\mathrm{in}| \hat{a}^{\mathrm{out}\,\alpha\,\dagger}(\omega)\,\hat{a}^{\mathrm{out}\,\alpha'}(\omega') |0^\mathrm{in} \rangle \nonumber \\ \nonumber&\quad= \!\!\!\!\!\!\!\sum\limits_{\substack{\beta, \beta' \notin \{ \alpha \},\{ \alpha' \}}}\!\!\!\!\!\!\! \mathcal{S}^*_{\alpha  \beta}(\omega) \,\mathcal{S}_{\alpha' \beta'}(\omega') \langle 0^\mathrm{in}| \hat{a}^{\mathrm{in}\,\beta}(\omega)\,\hat{a}^{\mathrm{in}\,\beta' \dagger}(\omega') |0^\mathrm{in} \rangle \\  & \quad =  \delta_{\{\alpha\}\{\alpha'\}}\, \delta(\omega-\omega')\,\sum\limits_{\beta \notin \{ \alpha \} } \mathcal{S}^*_{\alpha  \beta}(\omega) \,\mathcal{S}_{\alpha'  \beta}(\omega) .
\end{align}
 In \eqref{eq:2ndmom1} we have used \eqref{eq:Bogoljubov} and that the annihilation operator applied to the vacuum vanishes. In the second step we also used the commutator \eqref{eq:commutator}. The spectral correlation is $\delta$-function peaked as expected for a stationary process; there are no positive-to-negative norm correlations in the fields.

In what follows we will drop the explicit \textit{in}-vacuum state in the expectation value and the upper index \textit{in} on the operators. We also leave out the frequency dependence of $\mathcal{S}$ ($\hat{a}$), as it is corresponds with the first index of $\mathcal{S}$ (the mode of $\hat{a}$) in the moments calculation. Next, we calculate the normally ordered fourth order moment 
\begin{equation}
\label{eq:4thmom1}
\begin{split}
	\langle & \hat{a}^{\mathrm{out}\,\alpha\,\dagger}\,\hat{a}^{\mathrm{out}\,\alpha'\,\dagger}\,\hat{a}^{\mathrm{out}\,\alpha''}\,\hat{a}^{\mathrm{out}\,\alpha'''} \rangle \\  &= \!\!\!\!\!\!\sum\limits_{\substack{ \beta, \beta''' \notin \{ \alpha \},\{ \alpha''' \} }} \!\!\!\!\!\!\mathcal{S}^*_{\alpha  \beta} \,\mathcal{S}_{\alpha''' \beta'''} \langle \hat{a}^{\beta}\,\hat{a}^{\mathrm{out}\,\alpha'\,\dagger}\,\hat{a}^{\mathrm{out}\,\alpha''}\,\hat{a}^{\beta''' \dagger} \rangle \\  &= \!\!\!\!\!\!\sum\limits_{\substack{ \beta, \beta'''  \notin \{ \alpha \},\{ \alpha''' \}\\ \beta', \beta'' \in \{ \alpha' \},\{ \alpha'' \} }}\!\!\!\!\!\! \mathcal{S}^*_{\alpha  \beta} \,\mathcal{S}^*_{\alpha'  \beta'} \,\mathcal{S}_{\alpha''  \beta''} \,\mathcal{S}_{\alpha''' \beta'''} \langle \hat{a}^{\beta}\,\hat{a}^{\beta'\,\dagger}\,\hat{a}^{\beta''}\,\hat{a}^{\beta''' \dagger} \rangle  \\ &\quad \!\!+ \!\!\!\!\!\!\!\!\!\!\sum\limits_{\substack{ \beta, \beta' \notin \{ \alpha \},\{ \alpha' \}\\ \beta'', \beta''' \notin \{ \alpha'' \}, \{ \alpha''' \} }}\!\!\!\!\!\!\!\!\!\!\!\! \mathcal{S}^*_{\alpha  \beta} \,\mathcal{S}^*_{\alpha'  \beta'} \,\mathcal{S}_{\alpha''  \beta''} \,\mathcal{S}_{\alpha''' \beta'''} \langle \hat{a}^{\beta}\,\hat{a}^{\beta'}\,\hat{a}^{\beta''\,\dagger}\,\hat{a}^{\beta''' \dagger} \rangle,
	\end{split}
	\end{equation}
	with steps analogous to \eqref{eq:2ndmom1} and realizing that expectation values with unequal numbers of annihilation and creation \textit{in}-operators vanish. Next,
	\begin{equation}
	\label{eq:moments}
	\begin{split}
\!\!\!\!\langle \hat{a}^{\beta}\,\hat{a}^{\beta'\,\dagger}\,\hat{a}^{\beta''}\,\hat{a}^{\beta''' \dagger} \rangle&= \delta_{\beta \beta'} \, \delta_{\beta'' \beta'''} \,\delta(\omega-\omega')\,\delta(\omega''-\omega''')	\\ 
\!\!\!\!	\langle \hat{a}^{\beta}\,\hat{a}^{\beta'}\,\hat{a}^{\beta''\,\dagger}\,\hat{a}^{\beta''' \dagger} \rangle&= \delta_{\beta \beta''} \, \delta_{\beta' \beta'''} \,\delta(\omega-\omega'')\,\delta(\omega'-\omega''')\\&+\delta_{\beta \beta'''} \, \delta_{\beta' \beta''} \,\delta(\omega-\omega''')\,\delta(\omega'-\omega''),
	\end{split}
	\end{equation}
	due to the commutator. Inserting into \eqref{eq:4thmom1} and eliminating two sums with the Kronecker-deltas, we obtain the final expression
	\begin{widetext}
	\begin{equation}
	\label{eq:4thmom2}
	\begin{split}
		\langle &\hat{a}^{\mathrm{out}\,\alpha\,\dagger} \,\hat{a}^{\mathrm{out}\,\alpha'\,\dagger}\,\hat{a}^{\mathrm{out}\,\alpha''}\,\hat{a}^{\mathrm{out}\,\alpha'''} \rangle \\
&= \delta_{ \{\alpha\}\overline{\{\alpha'\}}} \, \delta_{\overline{\{\alpha''\}} \{\alpha'''\}} \,\delta(\omega-\omega')\,\delta(\omega''-\omega''') \sum\limits_{\substack{ \beta, \beta'' \notin \{ \alpha \},\{ \alpha''' \} }} \mathcal{S}^*_{\alpha  \beta} \,\mathcal{S}^*_{\alpha'  \beta} \,\mathcal{S}_{\alpha''  \beta''} \,\mathcal{S}_{\alpha''' \beta''} \\ 
&\quad+\delta_{\{\alpha\} \{\alpha''\}} \, \delta_{\{\alpha'\} \{\alpha'''\}} \,\delta(\omega-\omega'')\,\delta(\omega'-\omega''')\sum\limits_{\substack{ \beta, \beta' \notin \{ \alpha \},  \{ \alpha''' \} }} \mathcal{S}^*_{\alpha  \beta} \,\mathcal{S}^*_{\alpha'  \beta'} \,\mathcal{S}_{\alpha''  \beta} \,\mathcal{S}_{\alpha''' \beta'} \\ 
&\quad+ \delta_{\{\alpha\} \{\alpha'''\}} \, \delta_{\{\alpha'\} \{\alpha''\}} \,\delta(\omega-\omega''')\,\delta(\omega'-\omega'')\sum\limits_{\substack{  \beta, \beta' \notin \{ \alpha \},  \{ \alpha' \} }} \mathcal{S}^*_{\alpha  \beta} \,\mathcal{S}^*_{\alpha'  \beta'}\,\mathcal{S}_{\alpha''  \beta'} \,\mathcal{S}_{\alpha''' \beta}.
	\end{split}
\end{equation}	
\end{widetext}
In this expression we denote $\overline{\{\alpha\}}$ as the set of modes of norm opposite to that of mode $\alpha$. 
Finally, the not normally ordered fourth order moment of mode $\alpha$ is
\begin{widetext}
\begin{equation}
\label{eq:nnovariance}
\begin{split}
\langle  &\hat{a}^{\mathrm{out}\,\alpha\,\dagger}\,\hat{a}^{\mathrm{out}\,\alpha}\,\hat{a}^{\mathrm{out}\,\alpha\,\dagger}\,\hat{a}^{\mathrm{out}\,\alpha} \rangle \\ &= \langle  \hat{a}^{\mathrm{out}\,\alpha\,\dagger}\,\hat{a}^{\mathrm{out}\,\alpha\,\dagger}\,\hat{a}^{\mathrm{out}\,\alpha}\,\hat{a}^{\mathrm{out}\,\alpha} \rangle+\delta(\omega'-\omega'')\langle  \hat{a}^{\mathrm{out}\,\alpha\,\dagger}(\omega)\,\hat{a}^{\mathrm{out}\,\alpha}(\omega''') \rangle \\ & = \delta(\omega-\omega'')\,\delta(\omega'-\omega''')\sum\limits_{\substack{ \beta \notin \{ \alpha \} }} |\mathcal{S}_{\alpha  \beta}(\omega)|^{2} \sum\limits_{\substack{ \beta \notin \{ \alpha \} }} |\mathcal{S}_{\alpha  \beta}(\omega')|^2 
+ \delta(\omega-\omega''')\,\delta(\omega'-\omega'')\sum\limits_{\substack{ \beta \notin \{ \alpha \} }} |\mathcal{S}_{\alpha  \beta}(\omega)|^2 \sum\limits_{\substack{ \beta \notin \{ \alpha \} }} |\mathcal{S}_{\alpha  \beta}(\omega')|^2\\&\quad+ \delta(\omega'-\omega'')\,  \delta(\omega-\omega''')\sum\limits_{\beta \notin \{ \alpha \} } |\mathcal{S}_{\alpha  \beta}(\omega)|^2 ,
\end{split}
\end{equation}
\end{widetext}
which we obtain by applying the commutator \eqref{eq:commutator} in the first step and \eqref{eq:2ndmom1} and \eqref{eq:4thmom2} in the second step. The result leads to the variance of mode $\alpha$ \eqref{eq:variance}.

\end{document}